\documentclass[epj]{svjour}
\usepackage{graphics}
\RequirePackage{graphicx}
\RequirePackage{multirow}
\RequirePackage{yhmath}
\RequirePackage{hyperref}
\usepackage{color}
\usepackage{cases}
\usepackage{subfigure}
\usepackage{dcolumn,booktabs,bm}
\usepackage{slashed}
\usepackage{amsfonts,amssymb,stmaryrd,latexsym,amsmath}
\usepackage{textcomp}

\usepackage{array} 
\newcounter{RomanNumber}

\newcommand{\lyxmathsym}[1]{\ifmmode\begingroup\def\b@ld{bold}
  \text{\ifx\math@version\b@ld\bfseries\fi#1}\endgroup\else#1\fi}

\def\m{\mathcal}
\def\n{\nonumber}
\def\f{\frac}

\def\s{\sigma}

\def\a{\alpha}
\def\b{\beta}

\def\L{\Lambda}
\def\l{\lambda}

\def\X{\Xi_{cc}}
\begin{document}
\title{Possible hadronic molecules composed of the doubly charmed baryon and nucleon}
\author{Lu Meng\inst{1}\thanks{e-mail: lmeng@pku.edu.cn} \and Ning Li\inst{2}\thanks{e-mail: n.li@fz-juelich.de} \and Shi-Lin Zhu\inst{1,3}%
\thanks{e-mail: zhusl@pku.edu.cn} 
}                     

\institute{School of Physics and State Key Laboratory of Nuclear
Physics and Technology, Peking University, Beijing 100871, China\label{addr1} \and Institute~for~Advanced~Simulation,
Institut~f\"{u}r~Kernphysik, and
J\"{u}lich~Center~for~Hadron~Physics,~Forschungszentrum~J\"{u}lich,
D-52425~J\"{u}lich, Germany \label{addr2}\and  Collaborative Innovation Center of Quantum Matter, Beijing 100871, China \label{addr3}}
\date{Received: date / Revised version: date}
%
\abstract{
We perform a systematical investigation of the possible
deuteron-like bound states with configuration $\Xi_{cc}N (\bar{N})$,
where $N(\bar{N})$ denotes the nucleon (anti-nucleon), in the
framework of the one-boson-exchange-potential model. In the
spin-triplet sector we take into account both the ${}^3S_1$  and
${}^3D_1$ channels due to non-vanishing tensor force. There exist
several candidates of the loosely bound molecular states for the
$\Xi_{cc}N$ and $\Xi_{cc}\bar{N}$ systems, which lie below the
threshold of $\Lambda_c\Lambda_c$ or $\Lambda_c{\bar\Lambda}_c$. We also investigate the possible loosely bound states with configurations $\Lambda_cN(\bar{N})$ and $\Sigma_cN(\bar{N})$. These molecular candidates may be searched for at Belle II and LHC in
the near future.
%
} 
\maketitle
\section{INTRODUCTION}\label{Sec1}
Since the charmonium-like state $X(3872)$ was discovered by the
Belle Collaboration in 2003~\cite{Choi:2003ue}, more and more
charmonium-like/bottomonium-like states have been reported by the
experimental collaborations, such as
$Y(4260)$~\cite{Aubert:2005rm,He:2006kg}, $Y(10610)$,
$Y(10650)$\cite{Belle:2011aa,Abe:2007tk}, and
$Z_c(3900)$~\cite{Ablikim:2013mio,Liu:2013dau}, {\it et al}. Most
recently, the LHCb collaboration reported two hidden-charm
pentaquark states, $P_c(4380)$ and $P_c(4450)$, in
Ref.~\cite{Aaij:2015tga}. Such states are normally called ``exotic"
states since it is very difficult to interpret them using the
conventional hadron configurations, i.e., $q\bar{q}$ for a meson or
$qqq$ for a baryon in the quark model. Thus searching for the new
hadronic states beyond the quark model has become a hot topic. One
can refer to Refs. \cite{Chen:2016qju,Chen:2016spr,Guo:2017jvc} for recent
reviews on the experimental and theoretical progresses about the
exotic states in the past decades.

Since the masses of  many ``exotic" states are close to the
threshold of two hadrons, the hadronic molecule provides us an
appealing picture to interpret them. A hadronic molecule is a
loosely bound state formed by the color-singlet hadrons and the
force is generated by exchanging light bosons. Since the
meson-exchange force is some kind of residual force of the color
force generated by exchanging gluons between quarks, the hadronic
molecule may have a much larger size than the meson or baryon. For
example, the loosely bound deuteron is formed by the neutron and
proton and its root-mean-square radius is around $2$ fm. The
hadronic molecule composed of heavy mesons was first proposed by
Voloshin and Okun about forty years ago. They investigated the
possible molecular states formed by one charmed meson and one
charmed antimeson~\cite{Voloshin:1976ap}. De Rujula {\it et al}
tried to interpret $\psi(4040)$ as a $D^*\bar{D^*}$ molecular state
in Ref.~\cite{DeRujula:1976zlg}. T{\"o}rnqvist calculated the
possible deuteron-like $D\bar{D}^*$ and $D^*\bar{D}^*$ states in
Refs.~\cite{Tornqvist:1993ng,Tornqvist:1991ks}. Recently, many other
calculations were performed in the hadronic molecular picture, such
as the molecular states formed by two light
baryons~\cite{Straub:1990de,Huang:2004ke,Dai:2006gs,Zhang:2006dy,Chen:2007qn,Ping:2008tp,Huang:2011kf,Chen:2011zzb},
by two heavy
baryons~\cite{Lee:2011rka,Li:2012bt,Vijande:2016nzk,Carames:2015sya,Huang:2013rla,Gerasyuta:2011zx,Meng:2017fwb},
and by two heavy
mesons~\cite{Liu:2007bf,Liu:2008xz,Liu:2008tn,Liu:2009ei,Zhao:2015mga,Zhao:2014gqa,Meguro:2011nr,Froemel:2004ea}.

In 2002, the SELEX collaboration reported a doubly charmed baryon
with mass $3520$ MeV, named $\Xi_{cc}^{+}$, which contains two charm
quarks and one down quark~\cite{Mattson:2002vu}. Later this
structure was confirmed by the same collaboration
\cite{Ocherashvili:2004hi}. In the conference
report~\cite{Moinester:2002uw}, another doubly charmed baryon,
$\Xi_{cc}^{++}$, containing two charm quarks and one up quark was
reported at  $3780$ MeV. Most recently, the doubly charmed baryon
$\Xi_{cc}^{++}$  was observed by the LHCb collaboration at
$3621.40\pm0.72(\mbox{stat})\pm0.27(\mbox{syst})\pm0.14(\Lambda_c^+)$
MeV/$c^2$~\cite{Aaij:2017ueg}. Many theoretical work has been
performed to calculate the mass of the doubly charmed
baryons~\cite{Gershtein:1998sx,Itoh:2000um,Chang:2006eu,Zhang:2008rt,Wang:2010hs,Sun:2016wzh,Brodsky:2011zs,Chen:2017sbg,Li:2017cfz,Wang:2017mqp}.
With more and more information on $\Xi_{cc}$, it is very interesting
to investigate the structures formed by $\Xi_{cc}$ and other mesons
or baryons. In Refs.~\cite{JuliaDiaz:2004rf}, the authors studied
the hadronic molecule with configuration $\Xi_{cc}N$, and in
\cite{Meng:2017fwb} we investigated the possible deuteron-like bound
states of $\Xi_{cc}\Xi_{cc}(\bar{\Xi}_{cc})$. With the new mass
reported by LHCb, $\Xi_{cc}N(\bar{N})$ is about 13 MeV below the
threshold of $\Lambda_c\Lambda_c$, the possible $\Xi_{cc}N(\bar{N})$
molecular states do not decay into $\Lambda_c\Lambda_c ({\bar
\Lambda}_c)$ due to the limit phase space. Moreover, the
annihilation effect in the $\Xi_{cc}\bar{N}$ channel is strongly
suppressed compared with the nucleon anti-nucleon case since the
intermediate states are either the doubly charmed exotic teraquark
states or two charmed mesons which are rather heavy. In the current
work, we investigate the possible hadronic molecules with
configurations $B_cN(\bar{N})$ or $\Xi_{cc}N(\bar{N})$, where $B_c$
denotes $\Sigma_c$ or $\Lambda_c$ while $N(\bar{N})$ denotes the
nucleon (anti-nucleon).

We organize the paper as follows. We give the theoretical formalism
in Sec. \ref{Secform}, where the flavor wave functions, Lagrangians,
coupling constants and interaction potentials are presented. In Sec.
\ref{Secno}, we show the numerical results of the doubly charmed
baryon and nucleon systems and give some discussions. We summarize
our results in Sec.~\ref{Seccnclsn}. Finally, we collect some useful
formulae and the results of the singly charmed baryon and nucleon
systems in the Appendix.

\section{FORMALISM}\label{Secform}
In the present work, we
concentrate on the investigation of the possible molecular states
with configurations of $\Xi_{cc}N(\bar{N})$, $\Sigma_cN(\bar{N})$ and
$\Lambda_cN(\bar{N})$, where $\Lambda_c$ and $\Sigma_c$ is the spin-$1\over 2$ singly charmed baryons and $\Xi_{cc}$ is the spin-$1\over2$ doubly charmed baryon. Since we only consider the $u$ and $d$ quarks in the light quark sector,
the SU(2) symmetry should be applied. 

$\Lambda_c$  is the isospin singlet and $\Sigma_c$ is isospin triplet.  To be convenient, we construct
matrices as
\begin{eqnarray}
 \Sigma_{c}=\left(\begin{array}{cc}
\Sigma_{c}^{++} & \frac{1}{\sqrt{2}}\Sigma_{c}^{+}\\
\frac{1}{\sqrt{2}}\Sigma_{c}^{+} & \Sigma_{c}^{0}
\end{array}\right),\quad
\Lambda_{c}=\left(\begin{array}{cc}
0 & \Lambda_{c}^{+}\\
-\Lambda_{c}^{+} & 0
\end{array}\right).
\end{eqnarray}
The doubly charmed baryon is simpler since only one light quark is
involved. 
We adopt
$\Xi_{cc}=(\Xi_{cc}^u,~\Xi_{cc}^d)^T$ to denote the isospin-doublet
of the doubly charmed baryons, where the superscripts, $u$ and $d$,
denote the light quark in the baryon while ``$T$" means the
transpose of matrix. We also perform an investigation of the
possible molecular states with configuration of
$\Xi_{cc}N(\bar{N})$. We list the flavor wave functions in
Table~\ref{Table:Isospin}.
\begin{table*}[htp]
 \centering
 \caption{The flavor wave functions for the  systems with configurations of $\Sigma_{c}N$, $\Lambda_cN$ and
$\Xi_{cc}N$. $I$ denotes the isospin.}\label{Table:Isospin}
\begin{tabular}{lc|lc|lc}
\hline\hline
Systems          & Flavor wave functions                  & Systems            & Flavor wave functions            & Systems      & Flavor wave functions \\
\hline \multirow{4}{*}{$\left[\Sigma_{c}N\right]^{I=3/2}$} &
$\Sigma_{c}^{++}p$ &
\multirow{2}{*}{$\left[\Sigma_{c}N\right]^{I=1/2}$} &
$\sqrt{\frac{2}{3}}\Sigma_{c}^{++}n-\sqrt{\frac{1}{3}}\Sigma_{c}^{+}p$ & \multirow{3}{*}{$\left[\Xi_{cc}N\right]^{I=1}$} & $\Xi_{cc}^{u}p$ \\
                           & $\sqrt{\frac{1}{3}}\Sigma_{c}^{++}n+\sqrt{\frac{2}{3}}\Sigma_{c}^{+}p$ &       & $-\sqrt{\frac{2}{3}}\Sigma_{c}^{0}p+\sqrt{\frac{1}{3}}\Sigma_{c}^{+}n$ &
                           & $\frac{1}{\sqrt{2}}\left(\Xi_{cc}^{u}n+\Xi_{cc}^{d}p\right)$ \\
                           & $\sqrt{\frac{1}{3}}\Sigma_{c}^{\text{0}}p+\sqrt{\frac{2}{3}}\Sigma_{c}^{+}n$ &\multirow{2}{*}{$\left[\Lambda_{c}N\right]^{I=1/2}$} &
                              $\Lambda_{c}p$ &  & $\Xi_{cc}^{d}n$ \\
                           & $\Sigma_{c}^{0}n$ &  & $\Lambda_{c}n$ & $\left[\Xi_{cc}N\right]^{I=0}$ & $\frac{1}{\sqrt{2}}\left(\Xi_{cc}^{u}n-\Xi_{cc}^{d}p\right)$ \\
\hline \hline
\end{tabular}
\end{table*}

\subsection{The Lagrangian}\label{sec:lagrangian}
The Lagrangians for the  nucleon interacting with the light mesons
are,
\begin{eqnarray}
\mathcal{L}_{\pi NN}&=&\sqrt{2}g_{\pi NN}\bar{N}i\gamma_{5}\mathcal{M}_{\pi}N ,\\
\label{lagrangian:npi} \mathcal{L}_{\rho NN}&=&\sqrt{2}g_{\rho
NN}\bar{N}\gamma_{\mu}\mathcal{V}_{\rho}^{\mu}N 
+\frac{f_{\rho NN}}{\sqrt{2}m_{N}}\bar{N}\sigma_{\mu\nu}\partial^{\mu}\mathcal{V}_{\rho}^{\nu}N,\label{lagrangian:nrho}\\
 \mathcal{L}_{\omega NN}&=&\sqrt{2}g_{\omega
NN}\bar{N}\gamma_{\mu}\mathcal{V}_{\omega}^{\mu}N
+\frac{f_{\omega NN}}{\sqrt{2}m_{N}}\bar{N}\sigma_{\mu\nu}\partial^{\mu}\mathcal{V}_{\omega}^{\nu}N,\\
\label{lagrangian:nrho} \mathcal{L}_{\sigma NN}&=&g_{\sigma
NN}\bar{N}\sigma N, \label{lagrangian:nsigma}
\end{eqnarray}
where $N = (p, ~n)^T$ is the nucleon doublet. $m_N$ is the nucleon
mass and $g_{\pi NN}$, $g_{\rho NN}$, $f_{\rho NN}$, etc., are the
coupling constants, the values of which are given in
Tables~\ref{Table:mass} and \ref{Table:coupling}, respectively.
The light mesons are introduced with the  following notations,
\begin{eqnarray}
&&\mathcal{M_{\pi}}=\left(\begin{array}{cc}
\frac{\pi^{0}}{\sqrt{2}} & \pi^{+}\\
\pi^{-} & -\frac{\pi^{0}}{\sqrt{2}}
\end{array}\right),\\
&&\mathcal{V}_{\rho}^{\mu}=\left(\begin{array}{cc}
\frac{\rho^{0}}{\sqrt{2}} & \rho^{+}\\
\rho^{-} & -\frac{\rho^{0}}{\sqrt{2}}
\end{array}\right),\quad
\mathcal{V}_{\omega}^{\mu}=\left(\begin{array}{cc}
\frac{\omega}{\sqrt{2}} & 0\\
0 & \frac{\omega}{\sqrt{2}}
\end{array}\right),
\end{eqnarray}
where $\cal{M_{\pi}}$, $\mathcal{V}_{\rho}$ and
$\mathcal{V}_{\omega}$ are the $\pi$, $\rho$ and $\omega$ fields,
respectively. Following the Ref.~\cite{Riska:2000gd}, the interactions between heavy baryons and vector mesons can be introduced. Here we do not show the coupling  between the nucleon and $\eta$ because
it is rather small. Similarly, the Lagrangians for $\Xi_{cc}$
interacting with the light mesons read,
\begin{eqnarray}
\mathcal{L}_{\pi \Xi_{cc}\Xi_{cc}}&=&\sqrt{2}g_{\pi \Xi_{cc}\Xi_{cc}}\bar{\Xi}_{cc}i\gamma_{5}\mathcal{M}_{\pi}\Xi_{cc},\\
\mathcal{L}_{\rho \Xi_{cc}\Xi_{cc}}&=&\sqrt{2}g_{\rho
\Xi_{cc}\Xi_{cc}}\bar{\Xi}_{cc}\gamma_{\mu}\mathcal{V}_{\rho}^{\mu}\Xi_{cc}\nonumber \\
&& +\frac{f_{\rho \Xi_{cc}\Xi_{cc}}}{\sqrt{2}m_{\Xi_{cc}}}\bar{\Xi}_{cc}\sigma_{\mu\nu}\partial^{\mu}\mathcal{V}_{\rho}^{\nu}\Xi_{cc},\\
\mathcal{L}_{\omega \Xi_{cc}\Xi_{cc}}&=&\sqrt{2}g_{\omega
\Xi_{cc}\Xi_{cc}}\bar{\Xi}_{cc}\gamma_{\mu}\Xi_{cc}\mathcal{V}_{\omega}^{\mu}\nonumber \\
&&+\frac{f_{\omega \Xi_{cc}\Xi_{cc}}}{\sqrt{2}m_{\Xi_{cc}}}\bar{\Xi}_{cc}\sigma_{\mu\nu}\partial^{\mu}\mathcal{V}_{\omega}^{\nu}\Xi_{cc},\\
\mathcal{L}_{\text{\ensuremath{\sigma}}\Xi_{cc}\Xi_{cc}}&=&g_{\sigma
\Xi_{cc}\Xi_{cc}}\bar{\Xi}_{cc}\sigma \Xi_{cc}.
\end{eqnarray}
The masses of the doubly charmed baryons, $m_{\Xi_{cc}}$, and the
values of the coupling constants, $g_{\pi\Xi_{cc}\Xi_{cc}}$, etc.,
 are given in Tables \ref{Table:mass} and \ref{Table:coupling}, respectively.

For the interactions between $\Sigma_c$ ($\Lambda_c$), the corresponding Lagrangians read,
\begin{eqnarray}
\mathcal{L}_{\pi\Sigma_{c}\Sigma_{c}}&=&g_{\pi\Sigma_{c}\Sigma_{c}}\mbox{Tr}\left[\bar{\Sigma}_{c}i\gamma_{5}\mathcal{M_{\pi}}\Sigma_{c}\right], \\
\mathcal{L}_{\rho\Sigma_{c}\Sigma_{c}}&=&g_{\rho\Sigma_{c}\Sigma_{c}}\mbox{Tr}\left[\bar{\Sigma}_{c}\gamma_{\mu}\mathcal{V}_{\rho}^{\mu}\Sigma_{c}\right] \nonumber \\
& &+\frac{f_{\rho\Sigma_{c}\Sigma_{c}}}{2m_{\Sigma_{c}}}\mbox{Tr}\left[\bar{\Sigma}_{c}\sigma_{\mu\nu}\partial^{\mu}\mathcal{V}_{\rho}^{\nu}\Sigma_{c}\right],\\
\mathcal{L}_{\omega\Sigma_{c}\Sigma_{c}}&=&g_{\omega\Sigma_{c}\Sigma_{c}}\mbox{Tr}\left[\bar{\Sigma}_{c}\gamma_{\mu}\mathcal{V}_{\omega}^{\mu}\Sigma_{c}\right]\nonumber \\
&  &+\frac{f_{\omega\Sigma_{c}\Sigma_{c}}}{2m_{\Sigma_{c}}}\mbox{Tr}\left[\bar{\Sigma}_{c}\sigma_{\mu\nu}\partial^{\mu}\mathcal{V}_{\omega}^{\nu}\Sigma_{c}\right], \label{omega}\\
\mathcal{L}_{\sigma\Sigma_{c}\Sigma_{c}}&=&
g_{\sigma\Sigma_{c}\Sigma_{c}}\mbox{Tr}\left[\bar{\Sigma}_{c}\sigma\Sigma_{c}\right].\label{sigma}
\end{eqnarray}
For the isospin-singlet baryon $\Lambda_c$, it only interacts with
the light isospin-singlet mesons. Namely, there are only two
interaction vertices, $\Lambda_c \omega \Lambda_c$ and $\Lambda_c
\sigma \Lambda_c$. The corresponding Lagrangians are similar to
Eqs.~(\ref{omega}-\ref{sigma}).

\begin{table}[htp]
 \centering
 \caption{The masses of the one-charm baryon and light mesons are taken from \cite{Olive:2016xmw} while
 the mass for the doubly charmed baryon is taken from \cite{Aaij:2017ueg}. }\label{Table:mass}
\begin{tabular*}{0.5\textwidth}{@{\extracolsep{\fill}}lccc}
\hline\hline
Baryons & Mass (MeV) & Mesons & Mass (MeV)\\
\hline
Nucleon & 939 & $\pi$ & 137.27 \\
$\Sigma_{c}$ & 2454 & $\rho$ & 775.49 \\
$\Lambda_{c}$ & 2287 & $\omega$ & 782.65 \\
$\Xi_{cc}$ & 3621 & $\sigma$ & 600 \\
\hline \hline
\end{tabular*}
\end{table}

\subsection{Coupling Constants}\label{sec:coupling}
The coupling constants for the nucleons interacting with the light
mesons are well-known.  They are either extracted from the
experimental data or calculated by the theoretical models. We take
the values from
Refs.~\cite{Machleidt:2000ge,Cao:2010km,Machleidt:1987hj}. The other
coupling constants used in the current calculation can be derived
from those of the nucleons interacting with the light mesons through
the quark model. The same method as in Ref.~\cite{Meng:2017fwb} is
adopted. For the doubly charmed baryon case, we make use of the
following relations,
\begin{eqnarray}
\langle  p\uparrow |\mathcal{L}_{mNN}|p\uparrow \rangle &=& \langle p\uparrow |\mathcal{L}_{mqq}|p\uparrow \rangle, \label{cp:nucl}  \\
\langle \Xi_{cc}^u\uparrow |\mathcal{L}_{mhh}|\Xi_{cc}^u
\uparrow\rangle &=& \langle \Xi_{cc}^u\uparrow
|\mathcal{L}_{mqq}|\Xi_{cc}^u\uparrow \rangle.\label{cp:Bcc}
\end{eqnarray}
where ``$\uparrow$" means the third component of the spin is
$+\frac{1}{2}$. The matrix elements are calculated at both hadronic
and quark level. At the quark level, we adopt the following
Lagrangians~\cite{Riska:2000gd},
\begin{eqnarray}
\mathcal{L}_{q}&=&g_{\pi
qq}\bar{q}i\gamma_{5}\mathcal{M}_{\pi}q+g_{\rho
qq}\bar{q}\gamma_{\mu}\mathcal{V}_{\rho}^{\mu}q \nonumber \\
&& + g_{\omega
qq}\bar{q} \gamma_\mu \mathcal{V}_{\omega}^{\mu}q+g_{\sigma
qq}\bar{q}\sigma q
\end{eqnarray}
where $q=(u,d)^T$ is the light quark doublet. Here we adopt the non-relativistic chiral quark 
model proposed by Manohar and Georgi \cite{Manohar:1983md}, 
in which the pseudoscalar goldstone octet as well as the vector and 
scalar mesons are introduced as fundamental fields. Finally, we can obtain
the coupling constants used in the current calculation in terms of
those of the nucleon interacting with the light meson. The specific
expressions read,
\begin{itemize}
\item $\pi$-exchange,
\begin{eqnarray}
&&g_{\pi\Sigma_{c}\Sigma_{c}}=\frac{4\sqrt{2}}{5}\frac{m_{\Sigma_{c}}}{m_{N}}g_{\pi
NN}, \qquad g_{\pi\Lambda_{c}\Lambda_{c}}=0,\\\
&&g_{\pi\Xi_{cc}\Xi_{cc}}=-\frac{1}{5}\frac{m_{\Xi_{cc}}}{m_{N}}g_{\pi
NN}, \label{pi:exchange}
\end{eqnarray}
\item $\rho$-exchange,
\begin{align}
&g_{\rho\Sigma_{c}\Sigma_{c}}=2\sqrt{2}g_{\rho NN},\qquad g_{\rho\Xi_{cc}\Xi_{cc}}=g_{\rho NN},\\
&g_{\rho\Sigma_{c}\Sigma_{c}}+f_{\rho\Sigma_{c}\Sigma_{c}}=\frac{4\sqrt{2}}{5}\frac{m_{\Sigma_{c}}}{m_{N}}\left(g_{\text{\ensuremath{\rho}}NN}+f_{\text{\ensuremath{\rho}}NN}\right),\\
&g_{\rho\Xi_{cc}\Xi_{cc}}+f_{\rho\Xi_{cc}\Xi_{cc}}=-\frac{1}{5}\left(g_{\rho
NN}+f_{\rho NN}\right)\frac{m_{\Xi_{cc}}}{m_{N}},
\end{align}
\item $\omega$-exchange,
\begin{align}
&g_{\omega\Sigma_{c}\Sigma_{c}}=\frac{2\sqrt{2}}{3}g_{\omega NN},\qquad 
g_{\omega\Lambda_{c}\Lambda_{c}}=\frac{\sqrt{2}}{3}g_{\omega NN},
\\
&g_{\omega\Xi_{cc}\Xi_{cc}}=\frac{1}{3}g_{\omega NN},\qquad g_{\omega\Lambda_{c}\Lambda_{c}}+f_{\omega\Lambda_{c}\Lambda_{c}}=0,\\
&g_{\omega\Sigma_{c}\Sigma_{c}}+f_{\omega\Sigma_{c}\Sigma_{c}}=\frac{4\sqrt{2}}{3}\frac{m_{\Sigma_{c}}}{m_{N}}\left(g_{\omega NN}+f_{\text{\ensuremath{\omega}}NN}\right),\\
&g_{\omega\Xi_{cc}\Xi_{cc}}+f_{\omega\Xi_{cc}\Xi_{cc}}=-\frac{1}{3}\left(g_{\omega
NN}+f_{\omega NN}\right)\frac{m_{\Xi_{cc}}}{m_{N}},
\end{align}
\item $\sigma$-exchange,
\begin{eqnarray}
&&g_{\sigma\Sigma_{c}\Sigma_{c}} = \frac{2}{3}g_{\sigma NN}, \qquad
g_{\sigma\Lambda_{c}\Lambda_{c}} = \frac{1}{3}g_{\sigma NN},\\
&&g_{\sigma\Xi_{cc}\Xi_{cc}}    = \frac{1}{3}g_{\sigma
NN},\label{sigma:exchange}
\end{eqnarray}
\end{itemize}
the numerical values of which are given in
Table~\ref{Table:coupling}.
\begin{table}
 \centering
 \caption{The numerical values of the coupling constants used in the calculation.
 The values of the couplings for the nucleon interacting with the light mesons are taken from
 Refs.\cite{Machleidt:2000ge,Cao:2010km,Machleidt:1987hj}
 (Notice that for the $\omega NN$ coupling constant, a slightly larger value,
$16.7 - 23.1$ was determined in Ref.~\cite{Belushkin:2006qa}).
Others are calculated through Eqs.~(\ref{pi:exchange} -
\ref{sigma:exchange}).}\label{Table:coupling}
\begin{tabular*}{0.5\textwidth}{@{\extracolsep{\fill}}lcccccc}
\hline \hline
 $h$ & $g_{\pi hh}$& $g_{\sigma hh}$ & $g_{\rho hh}$ & $f_{\rho hh}$ & $g_{\omega hh}$ & $f_{\omega hh}$ \\
\hline
$N$ & 13.07 & 8.46 & 3.25 & 19.82 & 15.85 & 0.00 \\
$\Sigma_{c}$ & 38.65 & 5.64 & 9.19 & 59.01 & 14.95 & 63.17 \\
$\Lambda_{c}$ & 0.00 & 2.82 & 4.60 & -4.60 & 7.47 & -7.47 \\
$\Xi_{cc}$ & -10.08 & 2.82 & 3.25 & -21.04 & 5.28 & -25.67 \\
\hline \hline
\end{tabular*}
\end{table}

\subsection{The Interaction Potentials}
With the Lagrangians given in Sec.~\ref{sec:lagrangian}, we
calculate the scattering amplitude of the process $B+N\rightarrow
B+N$,  where $B$ and $N$ represent the heavy baryon and nucleon
respectively. The potentials have the following forms,
\begin{eqnarray}
V(\bm{Q}) &=& V_{cen}(\bm{Q}) + V_{SS}(\bm{Q}) \mathcal{O}_{SS} +V_{LS}({\bm Q}) \mathcal{O}_{LS} + V_{T}(\bm{Q}) \mathcal{O}_{T}\nonumber \\
\label{potential}
\end{eqnarray}
where $\mathcal{O}_{SS} = \bm{\sigma}_1 \cdot \bm{\sigma}_2$,
$\mathcal{O}_{LS}  = \frac{(\bm{\sigma}_1 + \bm{\sigma}_2)}{2} \cdot
(\bm{Q} \times \bm{k})$, and $\mathcal{O}_T = \bm{\sigma}_1 \cdot
\hat{Q} \bm{\sigma}_2 \cdot \hat{Q} - 3 \bm{\sigma}_1 \cdot
\bm{\sigma}_2$ are the spin-spin, spin-orbit, and tenor operators,
respectively.  $\bm{Q} = \bm{p}^\prime - \bm{p}$ is the momenta
transfer while $\bm{k} = \frac{\bm{p}^\prime + \bm{p}}{2}$ is the
averaged momenta between the incoming and outgoing nucleons. By
performing the Fourier transformation, $\bm{k}$ is replaced by $-i
\bm{\nabla}$ which provides the only nonlocal force in the current
calculation. Other nonlocal interactions, such as the recoil effect,
are very small and neglected. In the non-relativistic limit, we
expand $V(\bm {Q})$ as series of $\frac{\bm{Q}}{m_h}$ and truncate
at order $\mathcal{O}(\frac{\bm{Q}^2}{m_h^2})$, which coincides with
the Bonn model~\cite{Machleidt:1987hj}.

At each vertex, we introduce a monopole form factor
\begin{eqnarray}
\m{F}(\bm
Q)=\f{\L^2-m_{ex}^2}{\L^2-Q^2}=\f{\L^2-m_{ex}^2}{\l^2+\bm{Q}^2}\label{FF}
\end{eqnarray}
to suppress the high-momentum contribution. $\Lambda$ is the cutoff
parameter used to adjust the high-momentum contribution. $m_{ex}$ is
the exchanged meson mass and $\lambda^2=\Lambda^2-Q_0^2$.
Additionally, the form factor also accounts for the physics that the
constituent hadrons should be viewed as point-like particles in a
molecular state since they are well separated. Through the Fourier
transformation,
 \begin{eqnarray}
\m{V}(r)=\f{1}{(2\pi)^3}\int d\bm{Q} e^{i\bm{Q}\cdot\bm{r}}\m{V}(\bm
Q)\mathcal{F}^2(\bm Q),
\end{eqnarray}
we can obtain the interaction potentials in the coordinate space.
The specific expressions will be given later.

If the form factor were not introduced, there would be terms with a
delta function, $\delta^{(3)}(\bm{r})$, in the potentials after the
Fourier transformation. Such terms account for the very short
interaction and we call them contact interaction or delta
interaction. More details can be found in Appendix~\ref{app_func}.
In order to investigate the role of the delta interaction in the
formation of the loosely bound states, we make calculations both
with and without the delta interaction. Our potentials with the
delta interaction included in the coordinate space are,
\begin{itemize}
\item Pseudoscalar exchange:
\begin{eqnarray}
\m{V}_{SS}^p(r;\a)&=&C_{\a}^p\f{g_{1p}g_{2p}}{4\pi}\f{m_{\a}^3}{12M_1M_2}
H_1\bm{\s}_1\cdot\bm{\s}_2, \n\\
\m{V}_T^p(r;\a)&=&C_{\a}^p\f{g_{1p}g_{2p}}{4\pi}\f{m_{\a}^3}{12M_1M_2}
H_3S_{12}(\hat{r}), \label{potential:p1}
\end{eqnarray}
\item Vector exchange:
\begin{eqnarray}
\m{V}_C^v(r;\b)&=&C_{\b}^v\f{m_{\b}}{4\pi}\Big[g_{1v}g_{2v}H_0
+\f{m_{\b}^2}{16M_{1}^2}(g_{1v}g_{2v}+4g_{2v}f_{1v})H_1\n \\
&&~~~~~~~~~+ \f{m_{\b}^2}{16M_{2}^2}(g_{1v}g_{2v}+4g_{1v}f_{2v})H_1\Big], \n\\
\m{V}_{SS}^v(r;\b)&=&C_{\b}^v\left[g_{1v}g_{2v}+g_{1v}f_{2v}+g_{2v}f_{1v}+f_{1v}f_{2v}\right]\n \\
&&~~~~~~~~~\times\f{1}{4\pi}\f{m_{\b}^3}{6M_1M_2}H_1\bm{\s}_1\cdot\bm{\s}_2, \n\\
\m{V}_T^v(r;\b)&=&-C_{\b}^v\left[g_{1v}g_{2v}+g_{1v}f_{2v}+g_{2v}f_{1v}+f_{1v}f_{2v}\right]\n \\
&&~~~~~~~~~\times\f{1}{4\pi}\f{m_{\b}^3}{12M_1M_2}H_3S_{12}(\hat{r}), \n\\
\m{V}_{LS}^v(r;\b)&=&-C_{\b}^v\f{m_{\b}^3}{4\pi}H_2
\Biggl[\left(\frac{g_{1v}g_{2v}}{M_{1}M_{2}}\right)\bm{L}\cdot\bm{S}\n\\
&~&+\left(\frac{g_{1v}g_{2v}+2f_{1v}g_{2v}}{2M_{1}^{2}}+\frac{f_{1v}g_{2v}}{M_{1}M_{2}}\right)\bm{L}\cdot\bm{S}_1 \n\\
&~&+\left(\frac{g_{1v}g_{2v}+2f_{2v}g_{1v}}{2M_{2}^{2}}+\frac{f_{2v}g_{2v}}{M_{1}M_{2}}\right)\bm{L}\cdot\bm{S}_2
\Biggl], \label{potential:v}
\end{eqnarray}
\item Scalar exchange:
\begin{eqnarray}
\m{V}_C^s(r;\s)&=&-C_{\s}^sm_{\s}\f{g_{1s}g_{2s}}{4\pi}\large[H_0
-\f{m_{\s}^2}{16M_1^2}H_1-\f{m_{\s}^2}{16M_2^2}H_1\large],\n\\
\m{V}_{LS}^s(r;\s)&=&-C_{\s}^s\f{g_{1s}g_{2s}}{4\pi}\left(\f{m_{\s}^3}{4M_1^2}+\f{m_{\s}^3}{4M_2^2}\right)
H_2\bm{L}\cdot\bm{S}.\label{potential:s}
\end{eqnarray}
\end{itemize}
In the above expressions, the superscripts $p$, $s$ and $v$ denote
the pseudoscalar, scalar and vector mesons, respectively. $\alpha =
\pi$ while $\beta = \omega$ and $\rho$. $H_0$, $H_1$, $H_2$ and
$H_3$ are some functions with arguments, $r$, $\Lambda$ and
$m_{ex}$, refer to Appendix~\ref{app_func} for their specific
definitions. One can obtain the potentials without the short-range
delta interaction straightforwardly by making the simple
replacement, $H_1\rightarrow H_0$. $C_\alpha^p$, $C_\beta^v$ and
$C_\sigma^s$ are the isospin factors, the values of which are given
in Table~\ref{Table:isofactor}. The isospin-isospin factor is
already included in $C_\alpha^p$, etc.. That is why the operator
$\bm{\tau}_1 \cdot \bm{\tau}_2$ does not appear in
Eq.~(\ref{potential}). $\bm{L}\cdot \bm{S}(\bm{L} \cdot
\bm{S}_1,\bm{L}\cdot \bm{S}_2)$  is the spin-orbit operator which
provides the nonlocal force while $S_{12}(r) = \bm{\sigma}_1 \cdot
\hat{r}  \bm{\sigma}_2 \cdot \hat{r} - 3\bm{\sigma}_1 \cdot
\bm{\sigma}_2$ is the tensor force which has a non-vanishing matrix
element between the $S$ and $D$ waves. Thus, we also take into
account the $S-D$ mixing effect which plays a critical role in the
formation of the loosely bound deuteron. The matrix elements of the
operators are given in Appendix~\ref{app_matrix}.
\begin{table}[htp]
\centering \caption{Values of the isospin factors for $B_cN$ and
$\Xi_{cc}N$. The isospin factors  of the $B_c\bar{N}$ and
$\Xi_{cc}\bar{N}$ systems can be obtained easily by multiplying the
G-parity of the exchanged mesons. The G-parity of the exchange
mesons are given in the square brackets.}\label{Table:isofactor}
\begin{tabular*}{0.5\textwidth}{@{\extracolsep{\fill}}lcccc}
\hline\hline
Systems & $C_{\pi}^{p}$~[-1]  & $C_{\rho}^{v}$~[+1] & $C_{\omega}^{v}$~[-1] & $C_{\sigma}^{s}$~[+1] \\
\hline
$\left[\Sigma_{c}N\right]^{I=3/2}$ & $\frac{1}{\sqrt{2}}$ & $\frac{1}{\sqrt{2}}$ & $\frac{1}{\sqrt{2}}$ & $1$ \\
$\left[\Sigma_{c}N\right]^{I=1/2}$ & $-\sqrt{2}$ & $-\sqrt{2}$ & $\frac{1}{\sqrt{2}}$ & $1$ \\
$\left[\Lambda_{c}N\right]^{I=1/2}$ & $0$ & $0$ & $\sqrt{2}$ & $2$  \\
$\left[\Xi_{cc}N\right]^{I=1}$ & $1$ & $1$ & $1$ & $1$    \\
$\left[\Xi_{cc}N\right]^{I=0}$ & $-3$ & $-3$ & $1$ & $1$  \\
\hline\hline
\end{tabular*}
\end{table}

The spin of the systems composed of two spin-$\frac{1}{2}$ baryons
are $0$ (spin singlet) or $1$ (spin triplet). For the spin-singlet
case we focus on the ${}^1S_0$ channel while for the spin-triplet
case we take into account both the ${}^3S_1$ and ${}^3D_1$ channels
due to the existence of the tensor force. The wave function of the
${}^1S_0$ channel reads,
\begin{eqnarray}
\Psi(r,\theta,\phi)\chi_{ss_z}=y_S(r)|^1S_0\rangle,
\label{wave:function:1S0}
\end{eqnarray}
while that of the spin-triplet channel is
\begin{eqnarray}
\Psi(r,\theta,\phi)^T\chi_{ss_z}^T= \left(\begin{array}{c}
T_S(r) \\
 0   \\
\end{array}\right) |^3S_1\rangle +
\left(\begin{array}{c}
0  \\
T_D(r)\\
\end{array}\right)|^3D_1\rangle,\label{wave:function:3SD}\nonumber\\
\end{eqnarray}
where $y_S(r)$ is the radial wave function for the ${}^1S_0$ channel
while $T_S^T(r)$ and $T_D^T$ are the radial wave functions for the
${}^3S_1$ and ${}^3D_1$ channels respectively.

\section{Numerical results }\label{Secno}

We solve the Schr{\"o}dinger equation with the potentials derived in
the previous section and obtain the binding energy (B.E.) and the
radial wave function, with which we further calculate the
root-mean-square radius $r_{rms}$ . The binding energy and the
root-mean-square radius provide us the information to judge whether
a bound state exists. For the coupled channels, we additionally
calculate the individual probability for each channel.

A reasonable range of the cutoff in the study of the deuteron with
the OBEP model is $0.80 - 1.50$ GeV. Since the charmed baryon is
much heavier than the nucleon, we take a  wider range, $0.80 - 2.50$
GeV, for the cutoff parameter. 
 Notice that in Ref.~\cite{Ronchen:2012eg} an
even wider range, around $0.80 - 3.60$ GeV, was used for the cutoff
parameter and different values were used for different exchanged
mesons. 

The binding energy (B.E.) and root-mean-square radius ($r_{rms}$) are physical observables, 
from which we can judge if there are loosely bound states. The large B.E. and small $r_{rms}$ are not consistent with the molecular scheme. Recall the B.E. and  $r_{rms}$ of deuteron are $2.22$ MeV and $1.97$ fm~\cite{Machleidt:1987hj}, respectively, we choose $0.5<$B.E.$<15$ MeV and $0.8<r_{rms}<5$ fm as criteria for good candidates of molecular states.  In the numerical analysis, we list some results with B.E. and $r_{rms}$ beyond these ranges in order to see the cutoff dependence of our results.

We show the numerical results for the spin-singlet
$\Xi_{cc}N(\bar{N})$ system in Table~\ref{Table:BccNs0} and those
for the spin-triplet $\Xi_{cc}N(\bar{N})$ system in
Table~\ref{Table:BccNs1}.

\subsection{Spin-singlet (S = 0)}

\begin{table*}[htp]
 \centering
 \caption{The binding solutions for the spin-singlet $\Xi_{cc}N(\bar{N})$ systems. ``$\Lambda$" is the cutoff parameter.
 ``B.E." means the binding energy while $r_{rms}$ is the root-mean-square radius.}\label{Table:BccNs0}
\begin{tabular*}{0.8\textwidth}{@{\extracolsep{\fill}}lcccccc}
\hline \hline
                        & \multicolumn{3}{c}{With contact interaction} & \multicolumn{3}{c}{Without contact interaction} \\
               &   $\Lambda$ (GeV)  &   B.E (MeV)    &     $r_{rms}$ (fm)   & $\Lambda$ (GeV)   &    B.E (MeV) & $r_{rms}$ (fm) \\
\hline
\multirow{3}{*}{$[\Xi_{cc}N]^{(0,1)}$} & $1.28$ & $8.51$ & $1.17$ & $\times$ & $\times$ & $\times$ \\
                        &$1.30 $& $21.08$ & $0.98$ &             &               & \\
                        &$1.36$ & $88.30$ & $0.43$ &             &               & \\
\cline{2-7} \multirow{3}{*}{$[\Xi_{cc}N]^{(0,0)}$}
                      & $2.20$ & $0.49$ & $5.07$ & $1.00$ & $4.94$ & $2.06$  \\
                      & $2.40$ & $2.36$ & $2.95$ & $1.10$ & $25.75$ & $1.07$  \\
                      & $2.50$ & $5.67$ & $2.09$ & $1.20$ & $61.79$ &$ 0.78$ \\
 \hline
\multirow{3}{*}{$[\Xi_{cc}\bar{N}]^{(0,1)}$}
                      & $1.05 $& $1.98$ & $2.97$ & $1.05$ & $3.73$ & $2.26 $ \\
                      & $1.10$  & $9.78$ & $1.52$ & $1.10 $&$ 10.80 $& $1.46$ \\
                      & $1.20$ & $44.15$ & $0.85$ &$ 1.20$ & $35.18 $&$ 0.93$ \\
 \cline{2-7}
\multirow{3}{*}{$[\Xi_{cc}\bar{N}]^{(0,0)}$}
                     & $0.85$ & $4.98$ & $1.90$ & $0.90$ & $0.47$ & $4.73$ \\
                     & $0.90$ & $13.35$ & $1.31$ & $0.95$ & $24.73$ & $1.06$ \\
                     & $1.00 $& $34.86$ & $0.97$ &$ 1.00 $& $80.35 $& $0.71$\\
\hline \hline
\end{tabular*}
 \end{table*}

For the state $[\Xi_{cc}N]^{(0,1)}$, with the full potential we
obtain binding solutions which depend sensitively on the cutoff
parameter. After removing the delta interaction, the bound state
disappears. Thus, this state is not supported to be a candidate of
the hadronic molecule. For the state $[\Xi_{cc}N]^{(0,0)}$, we
obtain a loosely bound state with binding energy $ 0.49 <
\mbox{B.E.} <  10.39$ MeV for the cutoff to be $2.20 < \Lambda <
2.80$ GeV. Without the delta interaction, the binding energy becomes
$4.94 < \mbox{B.E.} < 61.79$ MeV for the cutoff to be $1.00 <
\Lambda < 1.20$ GeV. The root-mean-square radius is a few fm.
$[\Xi_{cc}N]^{(0,0)}$ can form a hadronic molecule with the
meson-exchange potential.

\begin{figure*}[htp]
\centering
\begin{tabular}{cc}
\includegraphics[width=0.45\textwidth]{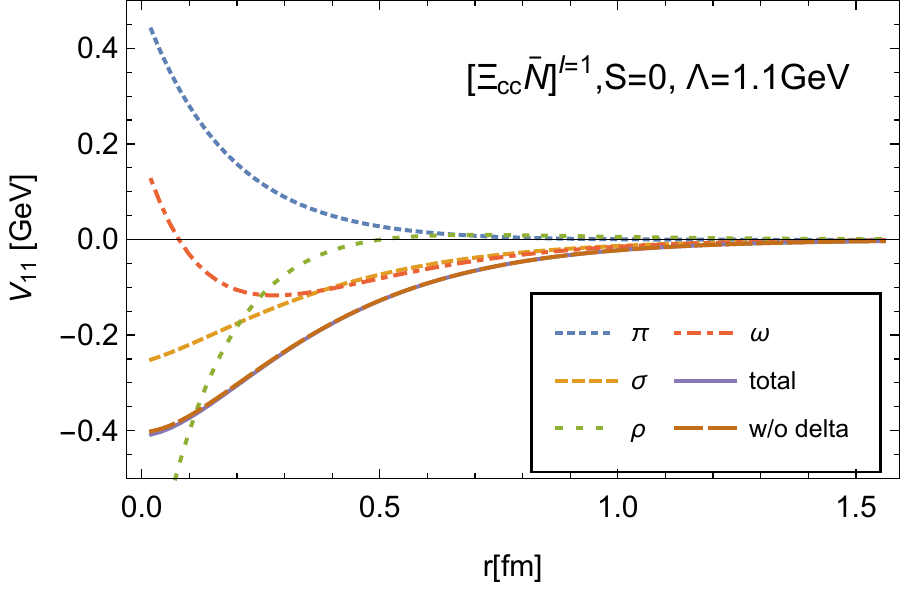} &
\includegraphics[width=0.45\textwidth]{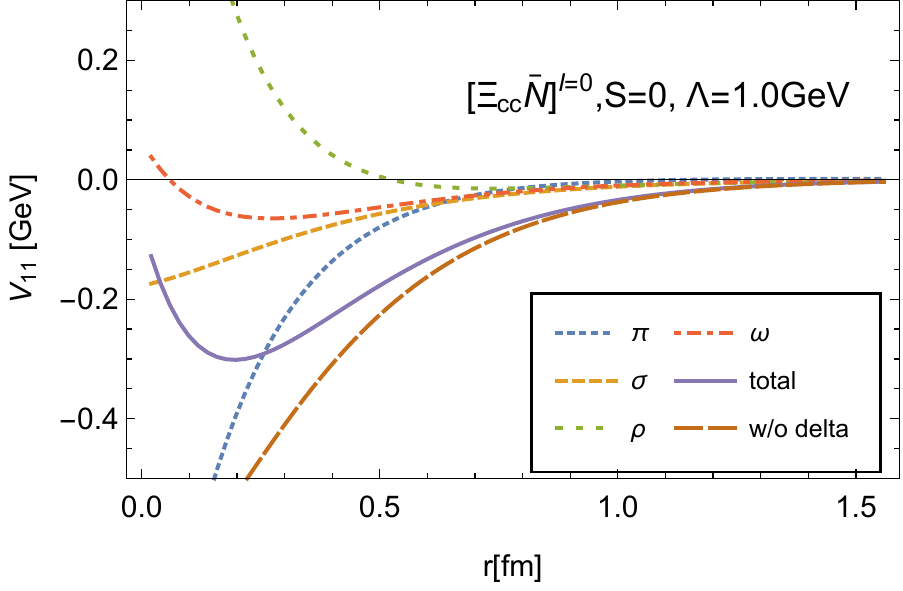}\\
\end{tabular}
\caption{The interaction potentials for the spin-singlet ($S = 0$)
$\Xi_{cc}\bar{N}$.
 ``w/o delta" means the delta interaction is removed from the total potential. }\label{pttlBccNs0}
\end{figure*}

The $\Xi_{cc}\bar{N}$ system is particularly interesting. We obtain
loosely bound states for both states with $(S,I) = (0, 0)$ and $(0,
1)$. We show the interaction potentials in Fig.~\ref{pttlBccNs0}.
From the plots, one can see clearly that both the $\sigma$- and
$\rho$-exchanges generate the attractive force for the state $(S, I)
= (0, 1)$ while both the $\sigma$- and $\pi$-exchanges provide the
attractive force for the states $(S, I) = (0, 0)$. In addition, the
$\omega$-exchange supplies the attractive force at long range for
both of the two states. For the state $[\Xi_{cc}\bar{N}]^{(0,0)}$,
the binding energy is $4.98 < \mbox{B.E.} < 34.86$ MeV for the
cutoff parameter to be $0.85 < \Lambda < 1.00$ GeV.  The
corresponding root-mean-square radius is $0.97 < r_{rms} < 1.90$ fm.
Thus, the system $[\Xi_{cc} \bar{N}]^{(0,0)}$ is favored to be a
candidate of the hadronic molecule. More interestingly, we obtain a
loosely bound state for $[\Xi_{cc}\bar{N}]^{(0,1)}$ both with and
without the delta interaction. The binding energy is $1.98 <
\mbox{B.E.} < 44.15$ MeV for the cutoff to be $1.05< \Lambda < 1.20$
GeV. With the same cutoff, the binding energy becomes $3.73 <
\mbox{B.E.} < 35.18$ MeV after neglecting the delta interaction. In
both cases, the root-mean-square radius is around $0.90 < r_{rms} <
2.5$ fm. The $[\Xi_{cc}\bar{N}]^{(0,1)}$ should be viewed as an
ideal candidate of the hadronic molecule.

\subsection{Spin-triplet (S = 1)}

For the state $[\Xi_{cc}N]^{(1,1)}$, we obtain no binding solutions.
Although we obtain bound state  for $[\Xi_{cc}N]^{(1,0)}$, the
results depend strongly on the cutoff parameter. Neither of the two
states is favored to be candidates of the loosely bound states.

\begin{figure*}[htp]
\centering
\begin{tabular}{cc}
\includegraphics[width=0.45\textwidth]{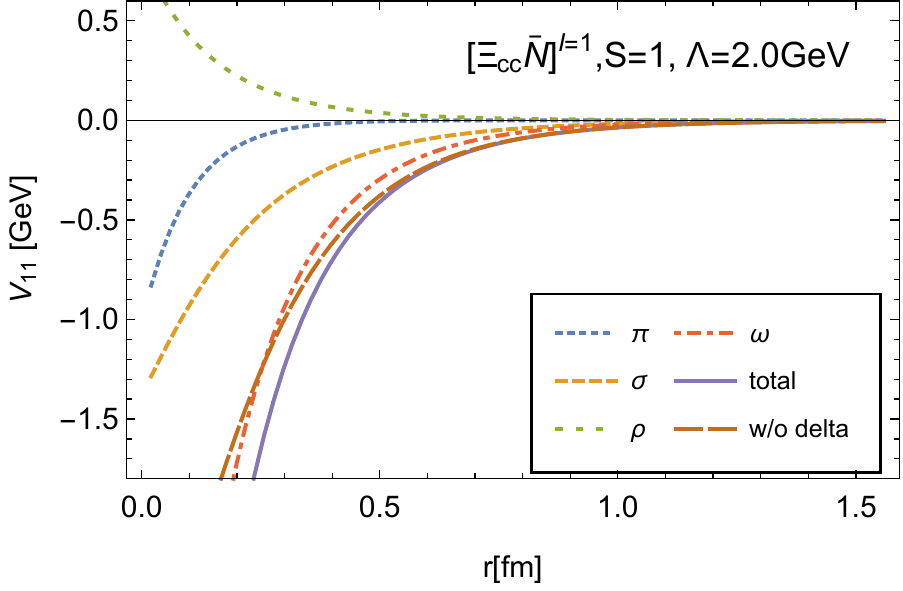}&
\includegraphics[width=0.45\textwidth]{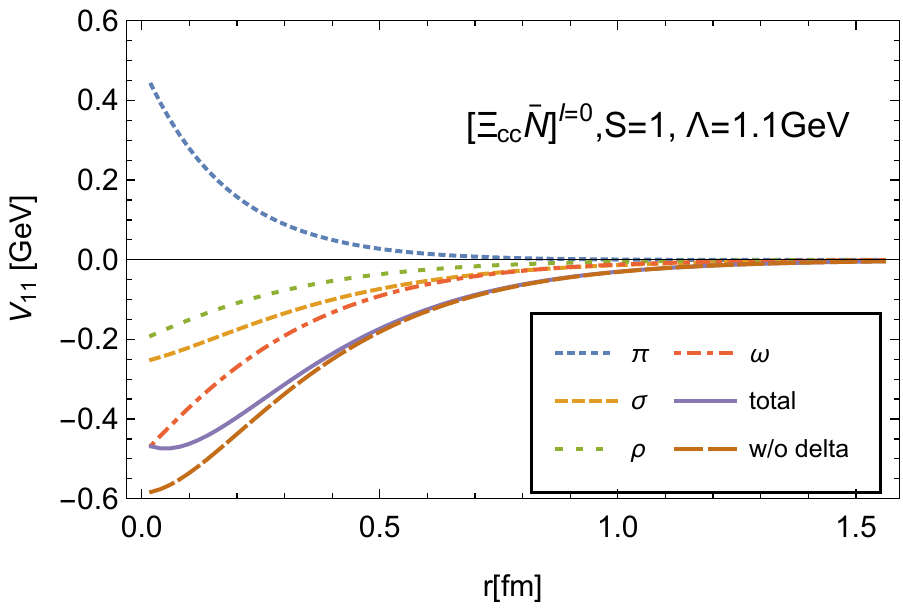}\\
\end{tabular}
\caption{The interaction potentials for the spin-triplet ($S = 1$)
$\Xi_{cc}\bar{N}$. Since the $S$ wave plays the dominant role, we
only plot the potential for the $^3S_1$ channel.
 ``w/o delta" means the delta interaction is removed from the total potential.}\label{pttlBccNs1}
\end{figure*}

Very interestingly, we obtain loosely bound states for
$[\Xi_{cc}\bar{N}]^{(1,0/1)}$, no matter the delta interaction is
included or not.  We show the interaction potentials in
Fig.~\ref{pttlBccNs1}. From the plots, one can see clearly that for
the state $(S, I) = (1, 0)$, the attractive force is generated by
the $\rho$-, $\omega$- and $\sigma$-exchanges while for the state
$(S, I) = (1, 1)$, the attractive force is provided by the $\pi$-,
$\omega$- and $\sigma$-exchanges.  With the full potential, the
binding energy of the state $[\Xi_{cc}\bar{N}]^{(1,0)}$ is $0.94 <
\mbox{B.E.} < 64.58$ MeV for the cutoff to be $0.95 < \Lambda <
1.10$ GeV. With the same cutoff, the binding energy becomes $4.81<
\mbox{B.E.} < 77.01$ MeV after one removes  the delta interaction.
For this state, the contribution of the $D$ wave is around $2\% -
5\%$, which is similar to the deuteron case, about $4\%$.  Based on
our results,  $[\Xi_{cc}\bar{N}]^{(1,0)}$ can be viewed a candidate
of the hadronic molecule. The state $[\Xi_{cc}\bar{N}]^{(1,1)}$ can
form an even more loosely bound state. The binding energy is $0.31 <
\mbox{B.E.} < 38.99$ MeV for the cutoff to be $0.95 < \Lambda <
1.10$ GeV. If one neglects the delta interaction, the binding energy
changes by a few MeV. The root-mean-square radius of this state is
about $0.90 - 4.5$ fm and the contribution of the $D$ wave is tiny,
around $0.2\% - 0.5\%$.  $[\Xi_{cc}\bar{N}]^{(1,1)}$ should also be
taken a ideal candidate of the hadronic molecule.

 \begin{table*}[htp]
 \centering
 \caption{The binding solutions for the spin-triplet ($S = 1$) $\Xi_{cc}N(\bar{N})$.
 ``$\Lambda$" is the cutoff parameter. ``B.E." means the binding energy while $r_{rms}$
 is the root-mean-square radius. $P_S$ is the probability (\%) of the S wave. }\label{Table:BccNs1}
\begin{tabular*}{0.9\textwidth}{@{\extracolsep{\fill}}lcccccccc}
\hline \hline
& \multicolumn{4}{c}{With contact interaction} & \multicolumn{4}{c}{Without contact interaction} \\
& $\Lambda$ (GeV) & B.E (MeV) & $r_{rms}$ (fm) & $P_{s}$ (\%) & $\Lambda$ (GeV) & B.E (MeV) & $r_{rms}$ (fm) & $P_{s}$ (\%) \\
\hline
$[\Xi_{cc}N]^{(1,1)}$ & $\times$ & $\times$ & $\times$ & $\times$ & $\times$ & $\times$ & $\times$ & $\times$ \\
\cline{2-9} \multirow{3}{*}{$[\Xi_{cc}N]^{(1,0)}$}
                &$ 1.97$ & $13.50 $&$ 0.64$ & $27.51$ &$ 2.00 $ & $ 7.13$ & $ 0.78 $ & $ 25.77$ \\
                 & $ 1.98 $ & $ 33.12 $ & $ 0.50 $ & $ 26.01  $ & $ 2.01 $ & $ 26.98 $ & $ 0.53 $ & $ 17.77$ \\
                & $1.99 $ & $ 54.83 $ & $ 0.48 $ & $ 18.14 $ & $ 2.02 $ & $ 49.04$ & $ 0.48 $ & $ 15.20$\\
 \hline
\multirow{3}{*}{$[\Xi_{cc}\bar{N}]^{(1,1)}$}
           &$ 0.95 $ & $ 0.31 $ & $ 5.31 $ & $ 99.85 $ & $ 1.00 $ & $ 0.71 $ & $ 4.33 $ & $ 99.79 $\\
           & $1.00 $ & $ 6.02  $ & $ 1.82 $ & $ 99.66 $ & $ 1.10 $ & $ 15.81 $ & $ 1.26 $ & $ 99.54$ \\
            & $ 1.10 $ & $ 38.99 $ & $ 0.89 $ & $ 99.59 $ & $ 1.20 $ & $ 48.30 $ & $ 0.83 $ & $ 99.47 $\\
\cline{2-9} \multirow{3}{*}{$[\Xi_{cc}\bar{N}]^{(1,0)}$}
       & $0.95$ & $ 0.94 $ & $4.05 $ & $ 98.48 $ & $ 0.95 $ & $ 4.81 $ & $ 2.09 $ & $ 97.67 $ \\
       & $1.00 $ & $ 10.17 $ & $ 1.57 $ & $ 96.77 $ & $ 1.00 $ & $ 18.59 $ & $ 1.24 $ & $ 96.66 $ \\
      & $ 1.10 $ & $ 64.58 $ & $ 0.80 $ & $ 94.88 $ & $ 1.10 $ & $ 77.01 $ & $ 0.75 $ & $ 95.03 $\\
\hline \hline
\end{tabular*}
\end{table*}

\begin{figure*}[htp]
\centering
\begin{tabular}{cc}
\includegraphics[width=0.45\textwidth]{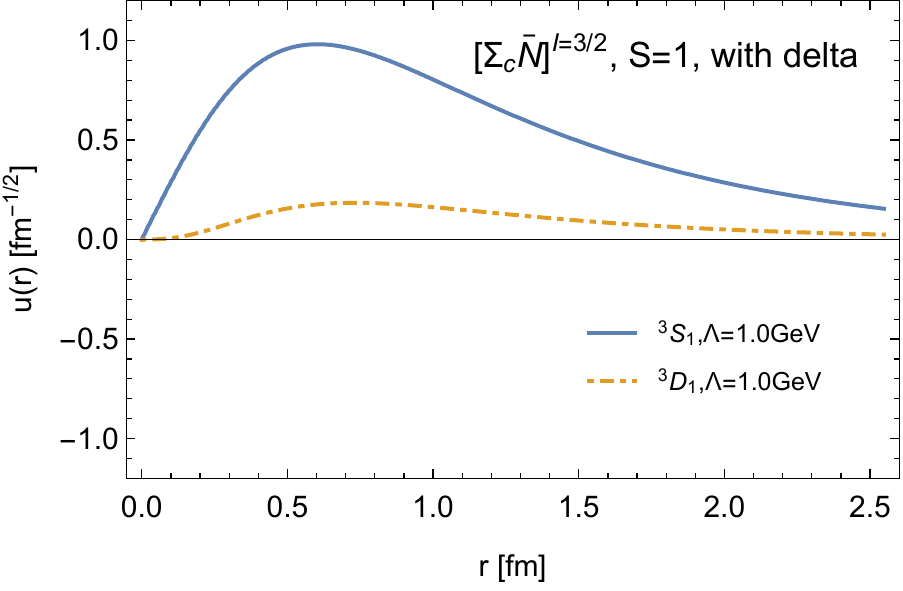}~&~
\includegraphics[width=0.45\textwidth]{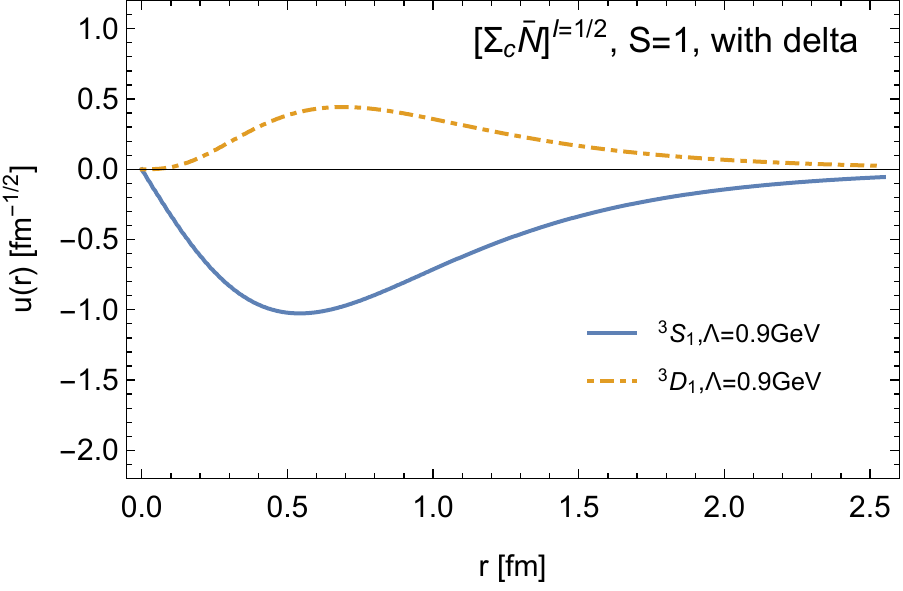}\\
\includegraphics[width=0.45\textwidth]{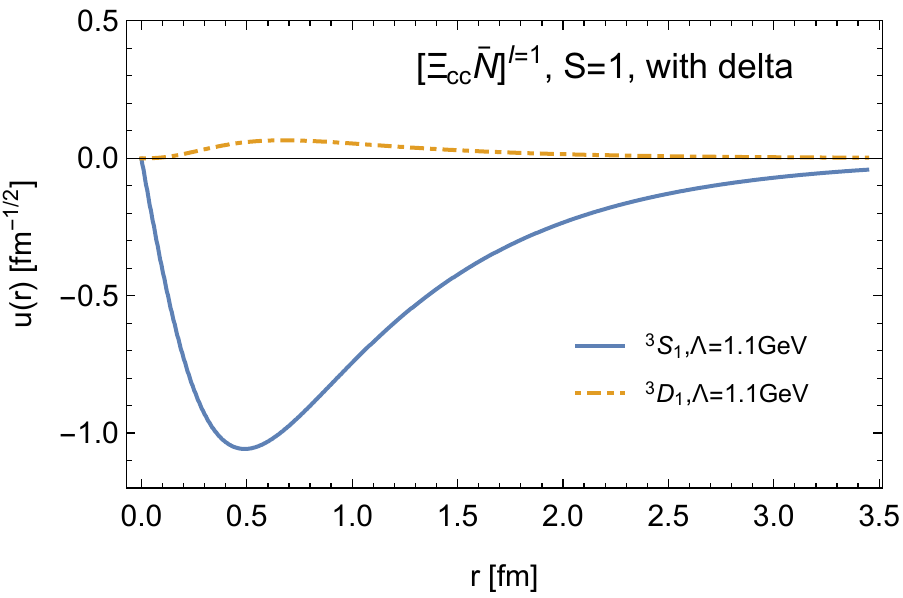}~&~
\includegraphics[width=0.45\textwidth]{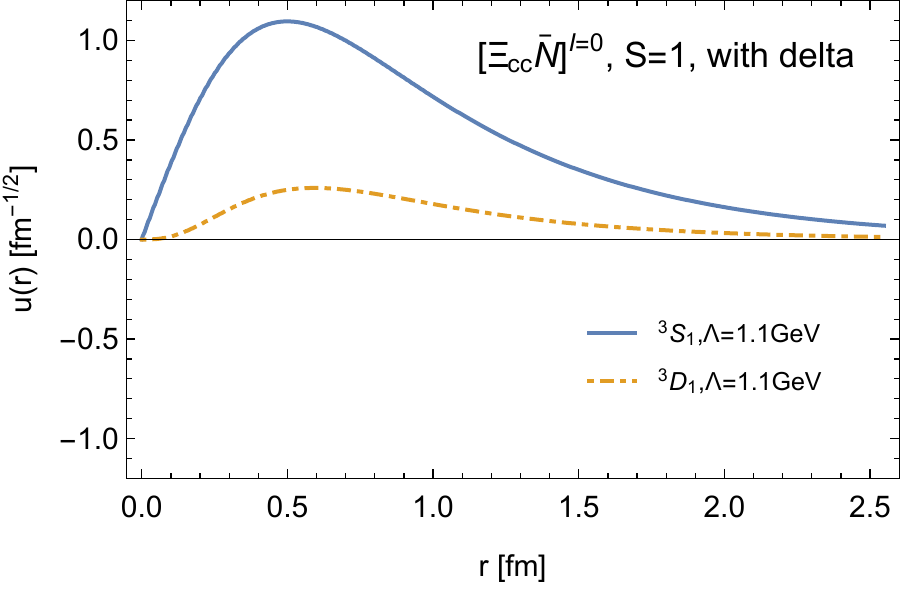}\\
\end{tabular}
\caption{The radial wave functions $u(r)=y(r)r$ for
states$[\Sigma_c\bar{N}]^{\left(1,\frac{1}{2}/\frac{3}{2}\right)}$
and $[\Xi_{cc}\bar{N}]^{(1,0/1)}$}\label{BcNur}
\end{figure*}

Finally, we show the radial wave functions for the state
$[\Xi_{cc}\bar{N}]^{(1,0/1)}$ in Fig.~\ref{BcNur}.

\section{DISCUSSIONS AND CONCLUSIONS}\label{Seccnclsn}

We have performed a systematic investigation of the possible
deuteron-like bound states with configurations $B_cN(\bar{N})$ and
$\Xi_{cc}N(\bar{N})$, where $B_c$ means $\Sigma_c$ (or $\Lambda_c$)
while $N (\bar{N})$ denotes the nucleon (anti-nucleon). In our
calculation, the one-boson-exchange potential model is applied.
Since in the hadronic molecule picture the constituent hadrons
should be taken as point-like particles, we apply one monopole form
factor for each vertex. By tuning the cutoff in the form factor one
can control the high-momentum contribution.  Additionally, to
investigate the the effect  of the very short-range interaction we
give the numerical results both with and without the delta
interaction.

For the spin-singlet ($S = 0$) case, we focus on the ${}^1S_0$
channel while in the spin-triplet  ($S = 1$) sector we take into
account both the ${}^3S_1$ and ${}^3D_1$ channels because of the
tensor force. Thus for the spin-triplet channels, a two-dimensional
coupled-channel Shr{\"o}dinger Equation is solved and the
probability of the $S$ wave is also given accordingly.

In the one-charm sector, without the coupled channel effects in the
flavor space we obtain bound states for neither spin-singlet or
spin-triplet $\Lambda_cN$. One can refer to Ref. \cite{Liu:2011xc}
for a coupled-channel analysis of this state. Our results indicate
that $[\Sigma_cN]^{\left(0,\frac{3}{2}\right)}$ and
$[\Sigma_cN]^{\left(1,\frac{1}{2}\right)}$ can be taken as the
candidates of the hadronic molecule. The $\Lambda_c\bar{N}$ system
is very interesting. Because of the vanishing tensor force for the
spin-triplet case, the spin-triplet states has the same numerical
results as those of the spin-singlet state. Both states can be
viewed as the candidates of the deuteron-like bound states.  The
$[\Sigma_c\bar{N}]$ system is also very interesting and all the spin
and isospin multiplets can form the loosely bound states by the
meson-exchange potentials. We also find that the delta interaction
has big influence on the $\Sigma_c\bar{N}$ system but it plays a
small role in the formation of the loosely bound $\Lambda_c\bar{N}$
state.


In the double-charm sector, none of the states
$[\Xi_{cc}N]^{(0,1)}$, $[\Xi_{cc}N]^{(1,0)}$,
$[\Xi_{cc}N]^{(1,1)}$ are supported to be the candidates of
the hadronic molecular states.  However, the state
$[\Xi_{cc}{N}]^{(0,0)}$ can form a loosely bound state with the
meson-exchange potentials. The $\Xi_{cc}\bar{N}$ is particularly
interesting. We obtain loosely bound states for all the four states:
$(S,I) = (0, 0)$, $(0,1)$, $(1,0)$ and $(1,1)$, with the
meson-exchange potentials. For the $[\Xi_{cc}\bar{N}]^{(0, 0/1)}$
systems, the delta interaction changes the binding energy by a few
to tens of MeV. The root-mean-square radius is around $0.8 - 2.5$
fm, which is comparable to that of the deuteron, about $2$ fm. For
the $[\Xi_{cc}\bar{N}]^{(1,0/1)}$ system, the delta interaction has
small effects, changing the binding energy by a few MeV. The
contribution of the $D$ wave to the state
$[\Xi_{cc}\bar{N}]^{(1,1)}$ is tiny, around $0.2\% - 0.5 \%$ while
for the $[\Xi_{cc}\bar{N}]^{(1,0)}$ system the probability of the
$D$ wave is about $2\% - 5\%$. All of the four states are supported
to be the candidates of the hadronic molecule by the
one-boson-exchange potential model.

 For the molecules composed of hadron and anti-hadron, there exit strong decay modes. The possible loosely bound states with configuration of $[B_c\bar{N}]$, can decay into one $D/D^*$ together with some photons
or light mesons. The $[\Xi_{cc}\bar{N}]$-type hadronic molecules may decay into two $D/D^*$s plus some photons or light mesons. 

For the molecules composed of a heavy baryon and nucleon, they can decay via the decay of their daughter 
particles. For example,  $[\Sigma_cN] \rightarrow \Lambda_cN\pi$, $\Lambda_c \rightarrow pK^{-}\pi^{+}$. Thus, the deuteron-like 
bound states of $[\Sigma_cN]$ can be searched by searching for its  invariant mass in the final states of  $NpK^{-}\pi^{+}\pi$. The strong decay of $[\Lambda_cN]$ is forbidden. Because the  weak decay model $\Lambda_c\rightarrow pK^-\pi^+$ takes the largest fraction ratio, the loosely bound state of $[\Lambda_cN]$ can also be searched by searching for its invariant mass in the final state of $NpK^{-}\pi^{+}$. The threshold of $\Xi_{cc}N$ is below that of $\Lambda_c\Lambda_c$, the strong decay of $[\Xi_{cc}N]$ is forbidden either. The  favorable decay modes is $[\Xi_{cc}N]^{(0,0)}\rightarrow \Lambda_cK^-\pi^+\pi^+p$~\cite{Aaij:2017ueg,Yu:2017zst}. Some of these possible deuteron-like bound states
may be searched for at BelleII and LHCb in the near future.

\bigskip\noindent\textbf{Acknowledgements:}
L. Meng is very grateful to G.J. Wang, H.S Li and B. Zhou for very
helpful discussions. The authors thank Ulf-G. Mei{\ss}ner for
helpful comments. This project is supported by the National Natural
Science Foundation of China under Grants NO. 11621131001, 11575008
and 973 program. This work is also supported in part by the DFG and
the NSFC through funds provided to the Sino-German CRC 110
``Symmetries and the Emergence of Structure in QCD".

\begin{appendix}

\section{Definitions of some functions and Fourier transform formulae} \label{app_func}

The definitions of the functions $H_i$ are \cite{Li:2012bt},
\begin{align}
H_0(\L,m,r)&=Y(u r)-\f{\l}{u}Y(\l r)-\f{r\beta^2}{2u}Y(\l r),\n \\ H_1(\L,m,r)&=Y(u r)-\f{\l}{u}Y(\l r)-\f{r\l^2\beta^2}{2u^3}Y(\l r), \n\\
H_2(\L,m,r)&=Z_1(u r)-\f{\l^3}{u^3}Z_1(\l r)-\f{\l
\beta^2}{2u^3}Y(\l r),\n\\
 H_3(\L,m,r)&=Z(u r)-\f{\l^3}{u^3}Z(\l
r)-\f{\l \beta^2}{2u^3}Z_2(\l r),
\end{align}
where,
\begin{eqnarray*}
 \beta^2=\L^2-m^2,\quad u^2=m^2-Q_0^2,\quad\l^2=\L^2-Q_0^2,
\end{eqnarray*}
and
\begin{eqnarray*}
 Y(x)&&=\f{e^{-x}}{x},\quad Z(x)=\left(1+\f{3}{x}+\f{3}{x^2}\right)Y(x),\n\\
 Z_1(x)&&=\left(\f{1}{x}+\f{1}{x^2}\right)Y(x),\quad Z_2(x)=(1+x)Y(x).
\end{eqnarray*}

In our case all heavy hadrons have the same masses, we have
\begin{eqnarray}
 Q_0^2 =\left(\sqrt{m_f^2+\bm{p}_f^2}-\sqrt{m_i^2+\bm{p}_i^2}\right)^2 \approx  {\left( \bm{p}_i+\bm{p}_f\right)^2\bm{Q}^2 \over {4m_{\X}^2}}.\n\\
 \end{eqnarray}
Thus $Q_0^2$ is a high-order term and can be directly dropped out.

Without the form factor, one makes Fourier transformation and
obtains
\begin{eqnarray}
&&\frac{1}{u^{2}+\bm{Q}^{2}}\rightarrow\frac{e^{-ur}}{4\pi r}=\frac{u}{4\pi}Y(ur), \\
&&\frac{\bm{Q}}{u^{2}+\bm{Q}^{2}}\rightarrow-i\nabla\left(\frac{u}{4\pi}Y(ur)\right)=i\frac{u^{3}}{4\pi}Z_{1}(ur)\bm{r},  \\
&&\frac{\bm{Q}^{2}}{u^{2}+\bm{Q}^{2}}\rightarrow-\frac{u^{3}}{4\pi}Y(ur)+\delta^{(3)}(\bm{r}), \label{ps} \\
&&\frac{Q_{i}Q_{j}}{u^{2}+\bm{Q}^{2}}\rightarrow-\frac{u^{3}}{12\pi}\left[Z(ur)k_{ij}+Y(ur)\delta_{ij}\right]+\frac{\delta_{ij}}{3}\delta^{(3)}(\bm{r}),\n\\\label{vector}
\end{eqnarray}
where $k_{ij}=3\f{r_ir_j}{r^2}-\delta_{ij}$. Clearly, there exist
terms with a delta function $\delta^{(3)}(\bm{r})$ in
Eqs.~(\ref{ps}-\ref{vector}). In the current work, we call these
terms the contact interaction or delta interaction.

After introducing the form factor, the Fourier transformation
formulae read
\begin{eqnarray}
&&\f{1}{u^2+\bm{Q}^2}\mathcal{F}^2(Q)\rightarrow\f{u}{4\pi}H_0(\L,m,r),\nonumber \\
&&\f{\bm{Q}^2}{u^2+\bm{Q}^2}\mathcal{F}^2(Q)\rightarrow-\f{u^3}{4\pi}H_1(\L,m,r), \nonumber \\
&&\f{\bm{Q}}{u^2+\bm{Q}^2}\mathcal{F}^2(Q)\rightarrow\f{iu^3}{4\pi}\bm{r}H_2(\L,m,r),\nonumber \\
&&\f{Q_iQ_j}{u^2+\bm{Q}^2}\mathcal{F}^2(Q)\rightarrow-\f{u^3}{12\pi}[H_3(\L,m,r)
k_{ij}\n\\
&&~~~~~~~~~~~~~~~~~~~~~~~~~~~~~~~~~~~+H_1(\L,m,r)\delta_{ij}]. \label{FTformula}
\end{eqnarray}
One can also get the results without the contact interaction term by
a simple replacement in the above equations,
\begin{equation}
H_1(\Lambda, m, r) \rightarrow H_0(\Lambda, m, r).
\end{equation}

\section{Matrix elements of the operators}\label{app_matrix}
In the present work, we encounter the following operators,
\begin{itemize}
\item Spin-spin operator:
\begin{equation}
\bm{\s}_1\cdot\bm{\s}_2,
\end{equation}

\item Spin-orbit operator:
\begin{equation}
\bm{L}\cdot\bm{S},\quad\bm{L}\cdot\bm{S}_1,
\quad\bm{L}\cdot\bm{S}_2,
\end{equation}

\item Tensor operator:
\begin{equation}
S_{12}(\hat{r})=3(\bm{\sigma}_1 \cdot \hat{r})(\bm{\sigma}_2 \cdot
\hat{r})-\bm{\sigma}_1 \cdot \bm{\sigma}_2.
\end{equation}

\end{itemize}
The derivation of matrix elements of these operators are given in
the Ref. \cite{Meng:2017fwb}. We give the results as follows,
\begin{itemize}
\item Spin-singlet ($S=0$):
\begin{eqnarray}
&&\bm{\s}_1\cdot\bm{\s}_2=-3,\quad S_{12}(\hat{r})=0,\n\\
&&\bm{L}\cdot\bm{S}=0,\quad\bm{L}\cdot\bm{S}_1=0,
\quad\bm{L}\cdot\bm{S}_2=0,
\label{operator:singlet}
\end{eqnarray}

\item Spin-triplet ($S=1$):
\begin{eqnarray}
&&\bm{\s}_1\cdot\bm{\s}_2 = \left(\begin{array}{cc}
 1  &   0  \\
 0  &   1  \\
 \end{array} \right),~~
 S_{12}(\hat{r})=
 \left(\begin{array}{ccc}
0   &  \sqrt{8}  \\
\sqrt{8}  &  -2   \\
\end{array}\right),\n\\
&&\bm{L}\cdot\bm{S}=2\bm{L}\cdot\bm{S_1} =2\bm{L}\cdot\bm{S_2}=\left(\begin{array}{cc}
0  &  0 \\
0  & -3 \\
\end{array}\right), \label{operator:triplet}
\end{eqnarray}
\end{itemize}

\section{The $B_cN$ and $B_c\bar{N}$ systems}

Throughout the work,  we use $B_c$ to denote the one-charm baryon,
$\Sigma_c$ or $\Lambda_c$. For the $\Sigma_cN(\bar{N})$ systems, the
$\pi$-, $\rho$-, $\omega$- and $\sigma$-exchanges are permitted
whereas only the $\omega$- and $\sigma$-exchanges are allowed for
the $\Lambda_cN(\bar{N})$ systems. We present the binding solutions
for the spin-singlet $B_cN(\bar N)$ systems in
Table~\ref{Table:BcNs0} and those for the spin-triplet
$B_cN(\bar{N})$ systems in Table~\ref{Table:BcNs1}.

\subsection{Spin-singlet (S = 0)}

For the spin-singlet ($S = 0 $) systems, we focus on the ground
state, namely the ${}^1S_0$ channel. The isospin of the
$\Lambda_cN(\bar{N})$ system is $\frac{1}{2}$ while that of the
$\Sigma_cN(\bar{N})$ system is either $\frac{1}{2}$ or
$\frac{3}{2}$. We denote the $\Lambda_cN$ system by
$[\Lambda_cN]^{(S,I)}$, where $S$ and $I$ denote the spin and
isospin respectively. The other systems are denoted in a similar
way. For the $[\Lambda_cN]^{\left(0/1,\frac{1}{2}\right)}$ system, a
systematic coupled-channel analysis has already been given in
Ref.~\cite{Liu:2011xc}. We present them again in the current work to
be self-consistent. Without the coupled-channel effects to the
$\Sigma_cN$($\Sigma_c^*N$) channels, one could not obtain binding
solutions. For the $[\Sigma_cN]^{\left(0,\frac{1}{2}\right)}$
system, the possible binding solutions depend strongly on the cutoff
and the delta interaction so we omit the results. For the
$[\Sigma_cN]^{\left(0,\frac{3}{2}\right)}$, we obtain no binding
solutions and a loosely bound state is produced if one removes the
delta interaction. This loosely bound state has binding energy $1.08
< \mbox{B.E.} < 34.82$ MeV for the cutoff to be $0.90 < \Lambda <
1.10$ GeV. The root-mean-square radius  is $1.01 < r_{rms} < 4.05$
fm, which indicates that it might be a candidate of the molecular
state.

\begin{figure*}[htp]
\centering
\begin{tabular}{cc}
\includegraphics[width=0.45\textwidth]{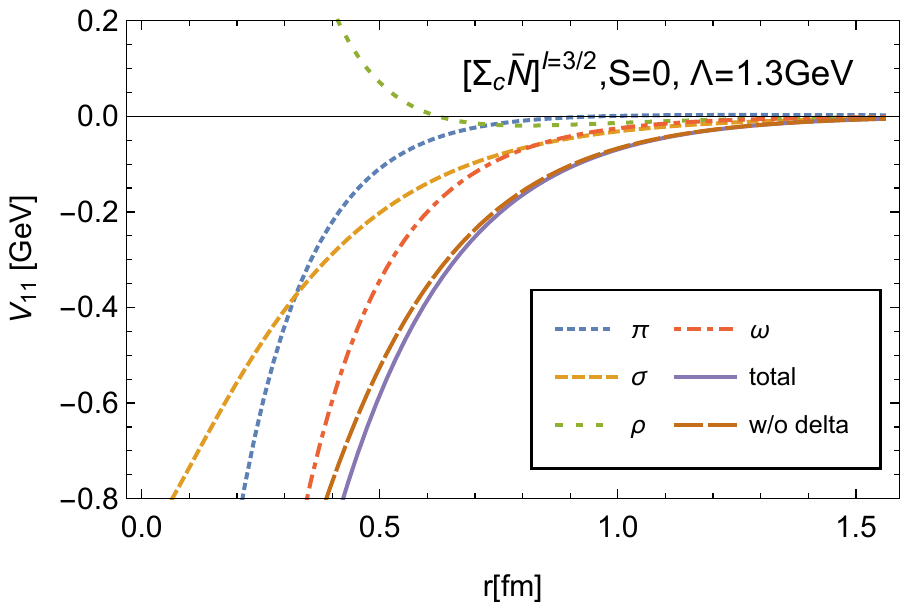} &
\includegraphics[width=0.45\textwidth]{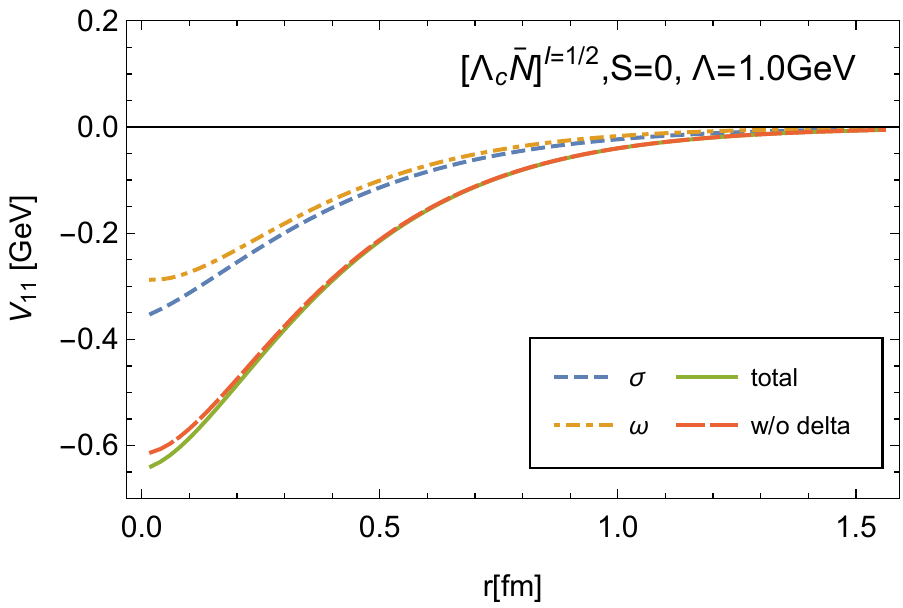}\\
\end{tabular}
\caption{The interaction potentials for the spin-singlet ($S = 0$)
$B_cN$ and $B_c\bar{N}$ systems.
 ``w/o" means the delta interaction is removed from the total potential.}\label{pttlBcNs0}
\end{figure*}

For the $[\Lambda_c\bar{N}]^{\left(0, \frac{1}{2}\right)}$ system,
only the $\omega$- and $\sigma$-exchanges are allowed by the
symmetry. The corresponding interaction potentials are shown in
Fig.~\ref{pttlBcNs0}, from which one can see clearly that both the
$\omega$ and $\sigma$-exchanges provide attractive force. Very
interestingly, we obtain a loosely bound state of
$[\Lambda_c\bar{N}]^{\left(0,\frac{1}{2}\right)}$, no matter the
delta interaction is included or not. We also find that the delta
interaction plays a minor role in the formation of the loosely bound
$[\Lambda_c\bar{N}]^{\left(0,\frac{1}{2}\right)}$, i.e., only
changing the binding energy by a few MeV.
 For example, the binding energy is $3.09 < \mbox{B.E.} < 56.51$ MeV for the cutoff to be $0.90 < \Lambda < 1.00$ GeV
with the total potential. With the same cutoff, the binding energy
is $2.51 < \mbox{B.E.} < 54.12$ MeV if one neglects the delta
interaction.  The $[\Lambda_c\bar{N}]^{\left(0,\frac{1}{2}\right)}$
should be taken as a promising candidate of the hadronic molecule.

For $[\Sigma_c\bar{N}]^{\left(0,\frac{1}{2}\right)}$, the possible
bound state is also very sensitive to the cutoff. After removing the
delta interaction, no binding solutions are obtained. Thus
$[\Sigma_c\bar{N}]^{\left(0, \frac{1}{2}\right)}$ is not a good
candidate of the molecular state.

The $[\Sigma_c\bar{N}]^{\left(0, \frac{3}{2}\right)}$ system is very
interesting. We obtain a loosely bound state both with and without
the delta interaction.  We show the interaction potentials in
Fig.~\ref{pttlBcNs0}, from which one can see clearly that the
$\pi$-, $\omega$- and $\sigma$-exchanges generate the attractive
force. With the full interaction, the binding energy is $10.78 <
\mbox{B.E.} < 70.82$ MeV for the cutoff to be $0.80 < \Lambda <
0.90$ GeV. If one neglects the delta interaction, the binding energy
is $0.56 < \mbox{B.E.} < 52.17$ MeV for the cutoff to be $0.90 <
\Lambda < 1.00$ GeV. The corresponding root-mean-square radius is
$0.73 < r_{rms} < 1.44$ fm  and $0.85 < r_{rms} < 4.63$ fm
respectively, which is comparable to that of the loosely bound
deuteron, around $2$ fm. The
$[\Sigma_c\bar{N}]^{\left(0,\frac{3}{2}\right)}$ system should be
taken as a candidate of the loosely bound state.

\begin{table*}[htp]
 \centering
 \caption{The binding solutions for the spin-singlet $B_cN$ and $B_c\bar{N}$ systems. ``$\Lambda$" is the cutoff parameter.
 ``B.E." means the binding energy while $r_{rms}$ is the root-mean-square radius.}\label{Table:BcNs0}
\begin{tabular*}{0.80\textwidth}{@{\extracolsep{\fill}}lcccccc}
\hline\hline
                      & \multicolumn{3}{c}{With contact interaction} & \multicolumn{3}{c}{Without contact interaction} \\
Systems        & $\Lambda$ (GeV) & B.E. (MeV) & $r_{rms}$ (fm) & $\Lambda$ (GeV) & B.E. (MeV) & $r_{rms}$ (fm) \\
\hline
\multirow{2}{*}{$\left[\Sigma_{c}N\right]^{\left(0,\frac{3}{2}\right)}$} & $\times$ & $\times$ & $\times$  &  $0.90$ & $1.08$ & $4.05$ \\
                     &             &                 &                  & $1.00$ & $12.06$ & $1.52$ \\
\hline
\multirow{3}{*}{$\left[\Sigma_{c}\bar{N}\right]^{\left(0,\frac{3}{2}\right)}$} &  $0.80$ & $10.78$ & $1.44$  & $0.90$ & $0.56$ & $4.63$ \\
 &$0.90$ & $70.82$ & $0.73$ & $1.00$ & $52.17$ & $0.85$ \\
 \cline{2-7}
\multirow{2}{*}{$\left[\Sigma_{c}\bar{N}\right]^{\left(0,\frac{1}{2}\right)}$} &$1.00$ & $1.85$ & $3.38$ & $\times$ & $\times$ & $\times$ \\
                     & $1.05$ & $58.06$  & $0.78$ &            &                & \\
 \cline{2-7}
\multirow{3}{*}{$\left[\Lambda_{c}\bar{N}\right]^{\left(0,\frac{1}{2}\right)}$} &$0.90$& $3.09$ & $2.58$ & $0.90$ & $2.51$ & $2.81$ \\
                    & $1.00$ & $56.51$ & $0.85$ & $1.00$ & $54.12$ & $0.86$ \\
                   \hline \hline
\end{tabular*}
\end{table*}

\subsection{Spin-triplet (S = 1)}

In the spin-triplet sector, both the ${}^3S_1$ and ${}^3D_1$
channels are considered because of the tensor force. Different from
the spin-singlet case, we additionally calculate the probability of
the $S$ wave. We present the binding solutions in
Table~\ref{Table:BcNs1}. Similar to the spin-singlet case, one can
refer to Ref. \cite{Liu:2011xc} for a systematical coupled-channel
analysis of $[\Lambda_cN]^{\left(1,\frac{1}{2}\right)}$. Without the
coupled-channel effects in the flavor space, one could not obtain
binding solutions. For the
$[\Sigma_cN]^{\left(1,\frac{1}{2}\right)}$ system, we could not find
binding solutions with the full interaction and obtain a loosely
bound state after removing the delta interaction from the total
potential. The binding energy is $1.57 < \mbox{B.E.} < 29.86$ MeV
with the cutoff parameter $0.80 < \Lambda < 1.20$ GeV. The
root-mean-square radius is $1.15 < r_{rms} < 3.62$ fm, which
indicates that $[\Sigma_cN]^{\left(1, \frac{1}{2}\right)}$ can form
a hadronic molecules with the meson-exchange potential. For this
state, the probability of the $D$ wave is round $5\%$. For the
system $[\Sigma_cN]^{\left(1,\frac{3}{2}\right)}$,  with the full
potential the binding solutions depend strongly on the cutoff
whereas no binding solutions are obtained  without the delta
interaction. Thus $[\Sigma_cN]^{\left(1,\frac{3}{2}\right)}$ is not
supported to be a candidate of the hadronic molecule.

\begin{figure*}[htp]
\centering
\begin{tabular}{cc}
\includegraphics[width=0.45\textwidth]{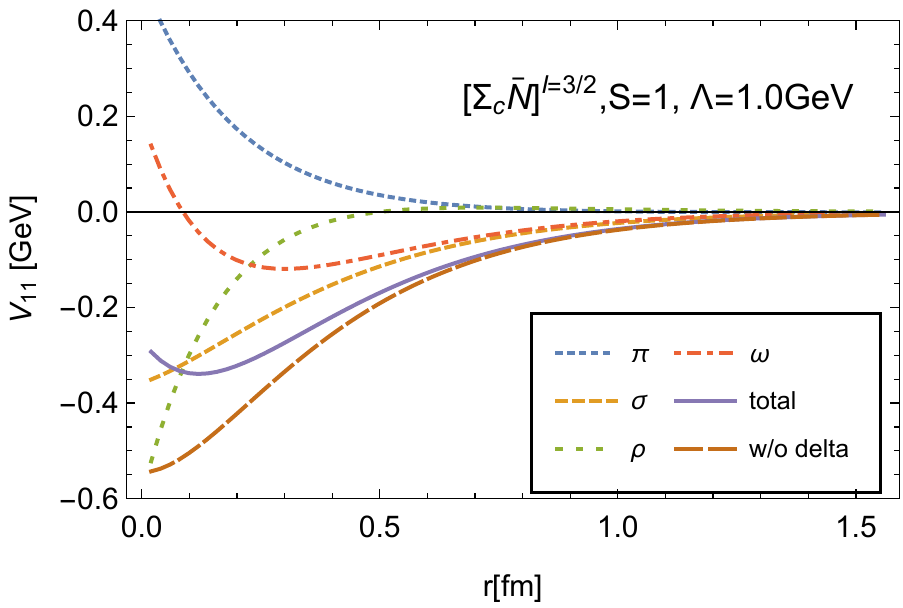} &
\includegraphics[width=0.45\textwidth]{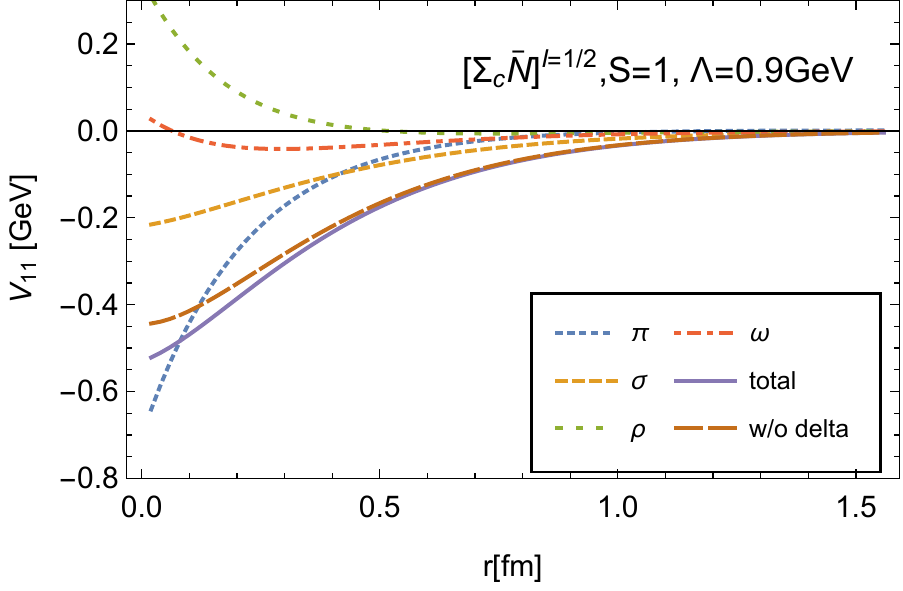} \\
\end{tabular}
\caption{The interaction potentials for
$[\Sigma_c\bar{N}]^{\left(1,\frac{1}{2}\right)}$ and
$[\Sigma_c\bar{N}]^{\left(1,\frac{3}{2}\right)}$. Since the $S$ wave
plays the dominant role, we only plot the interaction for the
${}^3S_1$ channel. ``w/o" means the delta interaction is removed
form the total potential. }\label{pttlBcNs1}
\end{figure*}

The results of $\Lambda_c\bar{N}$ do not depend on the spin because
the tensor force is vanishing for the spin-triplet case. The results
of $[\Lambda_c\bar{N}]^{\left(1,\frac{1}{2}\right)}$ are exactly the
same as those of $[\Lambda_c\bar{N}]^{\left(0,\frac{1}{2}\right)}$.
The $[\Sigma_c\bar{N}]$ system is very interesting. For both states
$(S,I) = (1,\frac{1}{2})$ and $(1,\frac{3}{2})$, we obtain loosely
bound states, no matter the delta interaction is considered or not.
We show the interaction potentials in Fig.~\ref{pttlBcNs1}. From the
plots, one can see clearly that the $\pi$- and $\sigma$-exchanges
generate the attractive force for
$[\Sigma_c\bar{N}]^{\left(1,\frac{1}{2}\right)}$ while the $\rho$-
and $\sigma$-exchange provide the attractive force for
$[\Sigma_c\bar{N}]^{\left(1,\frac{3}{2}\right)}$. The
$\omega$-exchange provides the repulsive force in the short range
and attractive force in the long range for both of the two states.
The contribution of the $D$ is around $10\% - 15\%$ for
$[\Sigma_c\bar{N}]^{\left(1,\frac{1}{2}\right)}$ and $2\% - 3\%$ for
$[\Sigma_c\bar{N}]^{\left(1,\frac{3}{2}\right)}$. Both of these two
systems can form hadronic molecules with the meson-exchange
potential.

\begin{table*}[htp]
 \centering
 \caption{The binding solutions for the spin-triplet $B_cN$ and $B_c\bar{N}$ systems. ``$\Lambda$"
 is the cutoff parameter. ``B.E." means the binding energy while $r_{rms}$ is the root-mean-square
 radius. $P_S$ is the probability (\%) of the S wave. }\label{Table:BcNs1}
\begin{tabular*}{0.9\textwidth}{@{\extracolsep{\fill}}lcccccccc}
\hline\hline
                     & \multicolumn{4}{c}{With contact interaction}       & \multicolumn{4}{c}{Without contact interaction} \\
Systems       &$\Lambda$ (GeV) & B.E (MeV) & $r_{rms}$(fm) & $P_{S}$ (\%) & $\Lambda$ (GeV) & B.E (MeV) & $r_{rms}$ (fm) & $P_{S}$ (\%) \\
\hline \multirow{3}{*}{$\left[\Sigma_{c}N\right]^{\left(1,
\frac{3}{2}\right)}$} &
                     $1.12$ & $3.70$ & $1.97$ & $99.47$ & $\times$ & $\times$ & $\times$ & $\times$ \\
                    &$1.15$& $17.51$ & $0.92$ & $99.81$ &           &        &          & \\
                    &$1.18$& $45.87$ & $0.60$ & $99.92$ &            &       &          & \\
\cline{2-9}
\multirow{3}{*}{$\left[\Sigma_{c}N\right]^{\left(1,\frac{1}{2}\right)}$}
&
                    $\times$ & $\times$ & $\times$ & $\times$ & 0.80 & 1.57 & 3.62 & 96.32 \\
                    &  &  &  &  & 1.00 & 17.85 & 1.40 & 93.94 \\
                     &  &  &  &  & 1.20 & 29.86 & 1.15 & 94.30 \\
\hline \multirow{2}{*}{$\left[\Sigma_{c}\bar{N}\right]^{\left(1,
\frac{3}{2}\right)}$} &
     0.90 & 1.17 & 3.91 & 98.51 & 0.90 & 6.24 & 1.98 & 97.85 \\
 & 1.00 & 36.70 & 1.03 & 96.91 & 1.00 & 54.76 & 0.88 & 97.13 \\
 \cline{2-9}
\multirow{2}{*}{$\left[\Sigma_{c}\bar{N}\right]^{\left(1,
\frac{1}{2}\right)}$ }
    & 0.80 & 17.31 & 1.39 & 90.54 & 0.80 & 1.66 & 3.43 & 92.27 \\
 & 0.90 & 88.98 & 0.83 & 85.11 & 0.90 & 81.77 & 0.86 & 84.63 \\
 \cline{2-9}
\multirow{2}{*}{$\left[\Lambda_{c}\bar{N}\right]^{\left(1,\frac{1}{2}\right)}$}
&
      0.90 & 3.09 & 2.58 & 100 & 0.90 & 2.51 & 2.81 & 100 \\
 & 1.00 & 56.51 & 0.85 & 100 & 1.00 & 54.12 & 0.86 & 100 \\
\hline \hline
\end{tabular*}
\end{table*}

In our calculation, we adopt the CD-Bonn potential model. Specifically, the $\pi$ exchange provides 
the long-range force, the $\sigma$ exchange supplies the medium-range potential and the heavier 
vector meson ($\rho$, $\omega$ and $\phi$) exchanges account for the short-range forces. In addition 
to the CD-Bonn one-boson-exchange potential (OBEP) model, there are also other OBEP approaches 
applied to the studies of the charmed-baryon-nucleon systems. For example, the Nijmegen group proposed 
an OBEP model with phenomenological hard-core potentials at short distance and applied it to the 
nucleon-nucleon and hyperon-nucleon systems \cite{Nagels:1975fb,Nagels:1976xq}. Following the Nijmegen 
OBEP approach, Dove and Kahana studied the possibility of the charmed baryons, $B_c(\Lambda_c, \Sigma_c, \Xi_c, \Xi_c^\prime)$, 
bind to a nucleon in \cite{Dover:1977jw},  Bahmathit analyzed the three-body system 
$B_cNN$ as well as $B_cN$ in \cite{Bhamathi:1981yu}, and Bando and Nagata investigated 
bindings of the $B_c-\alpha$ systems \cite{Bando:1983yt}. In these three calculations, the SU(3) 
symmetry used in the Nijmegen potential was extended to SU(4) in order to include the charmed baryons and 
mesons, although the charmed $D$ and $D^*$ exchanges are not important because of their heavy masses. 
Recently, Maeda {\it et al}. constructed a potential model (called ``CTNN") in which the long-range force arises from 
the $\pi$ and $\sigma$ exchanges while the short-range repulsion is evaluated by a quark cluster potential. 
Additionally, a monopole type form factor is introduced to the long-range potential to reflect the extended 
structure of hadrons \cite{Maeda:2015hxa}. We take the coupling constants $C_{\sigma}$ for $\sigma$ exchange and parameter $b$ for  Gaussian potential in the quark cluster model as two running parameters. In our calculations, we take four sets of parameters, 
\begin{itemize}
\item Set a:~~~~$ C_{\sigma}=-67.58,\quad b=0.6$fm;
\item Set b:~~~~$ C_{\sigma}=-77.50,\quad b=0.6$fm;
\item Set c:~~~~$ C_{\sigma}=-60.76,\quad b=0.5$fm;
\item Set d:~~~~$ C_{\sigma}=-70.68,\quad b=0.5$fm.
\end{itemize}
 \noindent A coupled-channel effect, 
$\Lambda_c N \leftrightarrow \Sigma_cN\leftrightarrow\Sigma_c^* N$, is also included in this calculation. 
We make a comparison of the present results with those of the previous studies, see Table \ref{comparison}.  
From the comparison, one can see clearly that the results of $\Lambda_c N$ are model-dependent in the 
``CTNN" approach even if the coupled-channel effect, $\Lambda_cN - \Sigma_cN -\Sigma_c^* N$, as well as the 
coulomb potential are included whereas the CD-Bonn OBEP model does not support $\Lambda_cN$ to form 
a bound state without coupling to $\Sigma_c N$ and $\Sigma_c^* N$. From our results, the system $[\Sigma_cN]^{(0,\frac{3}{2})}$ 
can be viewed as a candidate of the hadronic molecule, although the results depend slightly on how the delta potential 
is dealt with. This is consistent with the results from \cite{Dover:1977jw}.

\begin{table*}
\centering
\caption{Binding energies, in units of MeV,  of the charmed-baryon-nucleon systems with different OBEP approaches. ``$-$" denotes no relevant 
results while ``$\times$" means no bound states or the binding solutions is not stable at all.  
}\label{comparison}
\begin{tabular}{l|c|cccc|cc}
\hline\hline
\multirow{2}{*}{Channels}  & \multirow{2}{*}{ \cite{Dover:1977jw}}  &\multicolumn{4}{|c|}{CTNN \cite{Maeda:2015hxa}}  & \multicolumn{2}{c}{ this work}  \\
                                            &                           &       a            &               b        &          c                       &                d     &   with $\delta$  & without $\delta$   \\ 
\hline
$[\Lambda_cN]^{(0,\frac{1}{2})}$&   $-$           &   $  \times $   & $\times $        & $1.72\times10^{-3} $  &           $1.37 $ &   $\times$     & $\times $   \\  
$[\Lambda_cN]^{(1,\frac{1}{2})}$&   $-$           &  $  \times $    & $2.62 \times 10^{-4}$ & $1.97\times10^{-2}$ &  $1.57$ &   $\times$     & $\times $     \\              
$[\Sigma_cN]^{(0,\frac{1}{2})}$   & $-$	      &   $ - $            &   $ - $               &   $-$                           & $-$                  &   $ \times$    & $\times $   \\
$[\Sigma_cN]^{(0,\frac{3}{2})}$   & $1.76$       &   $-$              & $-$                   & $-$                             &  $ -$                &  $\times $     & $1.08 - 12.06$      \\
$[\Sigma_cN]^{(1,\frac{1}{2})}$   & $-$	      &   $- $             & $-$                   & $-$                             & $-$                  &  $\times$      & $1.57 - 29.86$    \\
$[\Sigma_cN]^{(1,\frac{3}{2})}$   & $-$	      &   $ -$             & $-$                   & $-$                             & $-$                  &  $3.70 - 45.87$  & $\times$   \\ 
\hline\hline
\end{tabular}
\end{table*}

\end{appendix}

\end{document}